\documentclass[aps,prl,reprint,nofootinbib,superscriptaddress]{revtex4-2}
\usepackage{graphicx}
\usepackage{dcolumn}
\usepackage{bm}
\usepackage{dsfont}
\usepackage{amsmath}
\usepackage{hyperref}
\usepackage{float}
\usepackage{caption}
\usepackage{natbib}
\usepackage{graphicx}
\usepackage{bbold}
\usepackage{physics}
\usepackage{mathtools}
\usepackage{MnSymbol}
\usepackage{relsize}
\usepackage{simpler-wick}
\usepackage{subfigure}
\usepackage{ragged2e}
\usepackage{pgfplots}
\pgfplotsset{compat=1.18}
\usepackage{xcolor}
\usepackage{natbib}

\renewcommand{\i}{\mathrm{i}}

\graphicspath{ {./imgs/} }
\DeclareGraphicsExtensions{.pdf,.png}
\usetikzlibrary{positioning}
\usepgfplotslibrary{fillbetween}

\begin{document}

\title{Collective enhancement in sideband cooling of ion crystals}
\author{Ivan Vybornyi}
\thanks{These authors contributed equally to this work.}
\affiliation{Institut für Theoretische Physik, Leibniz Universität Hannover, 30167 Hannover, Germany}
\author{Artem Zhdanov}
\thanks{These authors contributed equally to this work.}
\affiliation{Universität Innsbruck, Institut für Experimentalphysik, Technikerstraße 25, 6020 Innsbruck, Austria }
\author{Matthias Bock}
\affiliation{Universität Innsbruck, Institut für Experimentalphysik, Technikerstraße 25, 6020 Innsbruck, Austria }
\author{Leo Walz}
\affiliation{Universität Innsbruck, Institut für Experimentalphysik, Technikerstraße 25, 6020 Innsbruck, Austria }
\author{Nele Griesbach}
\affiliation{Universität Innsbruck, Institut für Experimentalphysik, Technikerstraße 25, 6020 Innsbruck, Austria }

\author{Klemens Hammerer}
\affiliation{Universität Innsbruck, Institut für Theoretische Physik, Technikerstraße 25, 6020 Innsbruck, Austria}
\affiliation{Institute for Quantum Optics and Quantum Information of the Austrian Academy of Sciences, 6020 Innsbruck, Austria}
\affiliation{Institut für Theoretische Physik, Leibniz Universität Hannover, 30167 Hannover, Germany}

\author{Christian F. Roos}
\affiliation{Universität Innsbruck, Institut für Experimentalphysik, Technikerstraße 25, 6020 Innsbruck, Austria }
\affiliation{Institute for Quantum Optics and Quantum Information of the Austrian Academy of Sciences, 6020 Innsbruck, Austria}

\date\today
\begin{abstract}
Low-entropy motional states of ion Coulomb crystals are an essential prerequisite for a plethora of applications and are typically prepared by laser cooling. As larger crystals are operated in the quantum regime, it remains unclear, and has recently become debated, whether increasing the ion number can be beneficial for cooling. Here, we investigate theoretically and experimentally many-ion sideband cooling and the role of collective effects in different spin--motion coupling regimes. For weak coupling, the many-body effects are insignificant. In the strong-coupling regime, however, the spin and motional subsystems undergo a coherent state swap, enabling cooling by a suitably timed laser pulse. Using planar Coulomb crystals with up to $91$ ions, we demonstrate that the residual mean phonon occupation after one such pulse scales as $1/N^2$ with the number of ions. By iterating the pulses, we measure mean phonon occupations $<2\cdot10^{-4}$. For large crystals in the coherent regime, we further show that the spin--motion dynamics becomes largely independent of the initial phonon statistics. Through spin measurements, the state-swap mechanism can be utilized to probe the phonon distribution in the mode. 
\end{abstract}

\maketitle

Coulomb crystals of trapped ions are among the leading platforms for quantum science and technology. Their high degree of control has enabled quantum computation with high-fidelity entangling gates
\cite{srinivas_high_fidelity_2021,zarantonello_robust_2019,loschnauer_scalable_2025},
programmable quantum simulation of complex many-body interactions \cite{guo_site_resolved_2024,monroe_programmable_2021,qiao_tunable_2024,joshi_observing_2022,franke_quantum_enhanced_2023}, and precision metrology, including timekeeping and tests of fundamental physics \cite{filzinger_multi_ion_2026,herschbach_linear_2012,hausser_2025,sokolik_direct_2026}. In all these directions, crystal size is crucial: scaling to larger Coulomb crystals enables more complex quantum simulations, larger quantum registers, and improved precision measurements. To date, experiments with crystals of more than one hundred trapped ions have been operated in the quantum regime \cite{jordan,kiesenhofer_controlling_2023}, with motional modes cooled close to the ground state. Increasing the crystal size poses new challenges, in particular for entropy management. These include enhanced heating \cite{jensen_temperature_2004,dubin_trapped_1999}, more demanding thermometry \cite{vybornyi_sideband_2023}, and a growing number of motional modes that must be cooled \cite{icc_exotic,chen_efficient-sideband-cooling_2020}.

Motional ground-state cooling of ion crystals is routinely achieved with established laser-cooling techniques. 
The underlying mechanism is the laser-induced coupling of the ions' electronic states to their motion, which transfers entropy from collective motional modes to internal degrees of freedom that are reset by optical dissipation. This mechanism is realized, for example, in resolved-sideband cooling \cite{diedrich_laser_1989,king_cooling_1998,monroe_resolved_sideband_1995,wu_continuous_2023} and in electromagnetically induced transparency (EIT) cooling \cite{morigi_ground_2000,morigi_cooling_2003,roos_experimental_2000}. 
In both cases, cooling has so far mainly been understood as a single-particle process, where each ion contributes independently to the cooling of a collective mode. However, in an ion crystal, laser-induced couplings to shared motional modes can mediate interactions among the electronic degrees of freedom, giving rise to the collective dynamics exploited in trapped-ion quantum simulation and computation. This raises the question whether laser cooling itself can benefit from collective many-body effects. For EIT cooling, the evidence for collective enhancement has been reported in Ref.~\cite{jordan} by comparing a many-ion experiment with a single-ion simulation. The effect appears to depend sensitively on the experimental configuration \cite{zhang_eit,qiao_eit,monroe} and the theory behind this has been recently explored in Ref.~\cite{khan_many_body_2026}. For sideband cooling, the corresponding question remains largely unexplored.

\begin{figure}[t]
\centering
\includegraphics{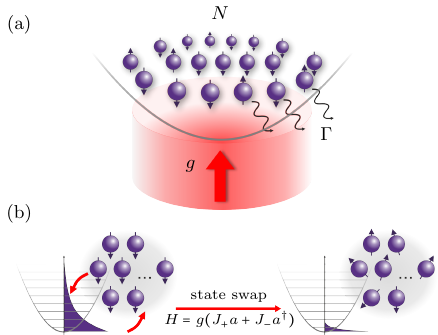}
\caption{\justifying (a) Schematic of the model. An arbitrary motional mode of a Coulomb crystal is coupled to the $N$ (effective) spins of ions via a global laser drive. (b) For large crystals and purely coherent interactions, the collective red-sideband pulse effectively swaps the states of the mode and the collective spin.} 
\label{fig:intro}
\end{figure}
We address this open question both theoretically and experimentally in a 2D trapped-ion simulator. We show that, for sideband cooling in the strong-coupling regime between the electronic states and a selected motional mode, suitably timed cooling pulses reduce the effective mode temperature by a factor scaling as $1/N^2$, with $N$ denoting the number of ions. This demonstrates a collective cooling enhancement, provided that $N$ is large as compared to the initial mean phonon number and most ions participate in the mode.
We derive this scaling theoretically and demonstrate it experimentally for selected center-of-mass (COM) and out-of-phase drumhead modes in planar ion crystals of different sizes, up to $N=91$ ions. The underlying mechanism is a coherent exchange of quantum states between the thermally occupied motional mode and the collective spin, initialized in a polarized state of effectively zero temperature, see Fig.~\ref{fig:intro}(b). Unlike pulsed sideband cooling based on driving a single ion \cite{wan_efficient_2015,rasmusson_optimized_2021,reed_comparison_2024,stutter_sideband_2018}, this collective swap is governed by the many-ion spin--motion coupling and therefore its characteristics are largely independent of the state of the mode. After the swap, the thermal excitations are carried by the spins and removed by optical pumping.

The same state-swap mechanism also enables an indirect measurement of the phonon statistics by counting spin excitations. This is relevant for estimating the mode occupation for non-thermal states, which typically requires more involved procedures \cite{rasmusson_high_2026,rasmusson_optimized_2021,chen_sympathetic_2017,hankin_systematic_2019}. While we derive and demonstrate these results for sideband cooling, we also show theoretically that suitable EIT-cooling configurations realize effectively the same collective dynamics. The role of strong coupling in collective enhancement of EIT cooling has recently been pointed out theoretically \cite{khan_many_body_2026}.

The state swap followed by optical pumping can be iterated until the lower limit is reached. In an experiment with an out-of-phase mode of a $19$-ion crystal we consistently observe mean occupations below $2\cdot 10^{-4}$, which, to the best of our knowledge, sets a new record for cooling of a mechanical oscillator. This value is expected to be set by the crystal heating and uncooled spectator modes, which limit both the cooling and the measurement via off-resonant coupling.

Here, we first introduce the many-ion cooling model and discuss the large-$N$ crossover between weak and strong spin--motion coupling. We then focus on the coherent pulsed regime, where collective spin--motion exchange leads to an enhanced cooling efficiency. Finally, we show, for different ion crystals and collective modes, that the cooling cycle can be iterated to reach exceptionally low mean phonon numbers.

We consider $N$ ions confined in a harmonic trap, forming an ion Coulomb crystal in the Lamb--Dicke regime, see Fig.~\ref{fig:intro}(a). Each ion has two relevant internal levels, $\ket{g}$ and $\ket{e}$, which define an effective spin-$1/2$. A laser field couples these internal states to a spectrally resolved collective vibrational mode of frequency $\nu$, described by annihilation and creation operators $a$ and $a^\dagger$. The spin--motion coupling strength is $g=\eta\Omega/2$, where $\eta\ll1$ is the single-ion Lamb--Dicke parameter and $\Omega$ is the carrier Rabi frequency. For a laser tuned to the red sideband, and neglecting the counter-rotating blue-sideband term in the rotating-wave approximation, $g\ll\nu$, the interaction Hamiltonian is
\begin{align}
    H=g(aJ_++ h.c.),
    \label{eq:rsb_main}
\end{align}
where $J_+=\sum_{i=1}^N\lambda_i\sigma^+_i$ is a mode-weighted collective spin ladder operator. The mode profile is encoded in normalized ion-dependent coefficients $\{\lambda_i\}_{i=1}^N$, with $\sum_{i=1}^N\lambda_i^2=1$. Cooling is completed by incoherent relaxation on the driven internal transition. We assume that each ion decays independently from $\ket{e}$ to $\ket{g}$ with rate $\Gamma$, thereby resetting the internal degrees of freedom and providing an entropy sink for the motional mode. The joint dynamics of the selected mode and the spins is then described by the master equation
\begin{align}
   \dot{\rho}=-i[H,\rho]+\Gamma\sum_{i=1}^N\mathcal{D}[\sigma_i^-]\rho ,  \label{eq:meq}
\end{align}
with $\mathcal{D}[L]\rho=L\rho L^\dagger-\frac{1}{2}\left\{L^\dagger L,\rho\right\}.$ This model covers experimental implementations of sideband cooling with narrowline one-photon and Raman transitions and can also be viewed as an effective description of a particular EIT configuration (similar to the one treated in Ref.\cite{shankar,jordan}, see Supplemental Material \cite{supplemental}). 

For continuous driving in the weak-coupling regime, $g\ll\Gamma$, the driven transition produces continuous damping of the selected motional mode. For a single ion, $N=1$ and $\lambda_1=1$, the master equation implies an effective cooling rate $\gamma_1=4g^2/\Gamma$~\cite{leibfried_quantum_2003}. For an ion crystal, the ions contribute independently in this regime, so that $\gamma_N=4\sum_i(g\lambda_i)^2/\Gamma=\gamma_1$. Hence, the increased number of ions is compensated by the reduced coupling of each ion to the normalized collective mode, and there is no collective enhancement of the cooling rate.

To discuss the crossover from weak to strong coupling, it is instructive to first consider the COM mode, for which $\lambda_i=1/\sqrt{N}$. In the large-$N$ limit, the master equation yields the following time dependence of the mean phonon number in the mode,
\begin{align}
    \bar{n}=\cramped{\dfrac{\bar{n}_0e^{-\frac{\tau}{2}}}{2\omega^2}
    \left[
    \omega^2-1
    +(1+\omega^2)\cosh\frac{\omega\tau}{2} 
    +2\omega\sinh\frac{\omega\tau}{2}
    \right]},
    \label{eq:nb_an}
\end{align}
as shown in the Supplemental Material \cite{supplemental}. Here, $\omega=\sqrt{1-(4g/\Gamma)^2}$, $\tau=\Gamma t$, and $\bar{n}_0$ is the initial mean phonon number. In the weak-coupling (overdamped) regime, Eq.~\eqref{eq:nb_an} reduces to an exponential decay with the single-ion cooling rate $\gamma_1$, independent of $N$, in agreement with the discussion above. In the strong-coupling, or underdamped, regime, $g\gg\Gamma$, the dynamics becomes coherent and approximately follows $\bar{n}\simeq \bar{n}_0 e^{-\Gamma t/2}\cos^2(gt)$. 
Although the envelope itself does not depend on $N$, collective enhancement can be observed in the coherent transient dynamics of the underdamped regime.
The oscillations of $\bar n(t)$ produce near-zeroes of the motional occupation at well-defined times $gt\approx\pi/2$ where the motional excitations are collectively transferred to the spins. Stopping the sideband drive at this time and subsequently reinitializing the spins by means of optical pumping realizes a pulsed cooling step. As shown below, the exact residual mode excitation at this point decreases with the number of ions, and thus proves collective cooling enhancement. This mechanism applies not only to the COM mode but also to non-COM modes, given the crystal is sufficiently large for the collective large-$N$ dynamics to emerge and the majority of the ions participate in the mode with comparable coupling strength. 

To study this phenomenon experimentally, we prepare a planar ion crystal of $^{40}$Ca$^+$ ions in a monolithic Paul trap~\cite{kiesenhofer_controlling_2023}. The electronic states of the ions are initialized in $4S_{1/2}, m_J=-1/2$ via optical pumping. To suppress spurious signals from spectator modes, all out-of-plane  modes of the ion crystal are laser-cooled to near their ground state by EIT cooling \cite{kiesenhofer_controlling_2023}. Following that, the highest-frequency non-COM drumhead mode of the crystal is prepared in a thermal state of about $\bar n_0\approx6$ via continuous sideband heating on the blue sideband of the narrow 729\,nm $4S_{1/2}\rightarrow3D_{5/2}$ transition. The choice of this mode over the COM mode is motivated by the significantly lower heating rate of non-COM modes in a macroscopic ion trap ($0.5$ ph/s/ion for the COM mode and negligible heating of the other modes at short times). 
The spin-motion coupling is induced by driving the red sideband of the same electronic transition.
\begin{figure}[t]
\centering
\includegraphics{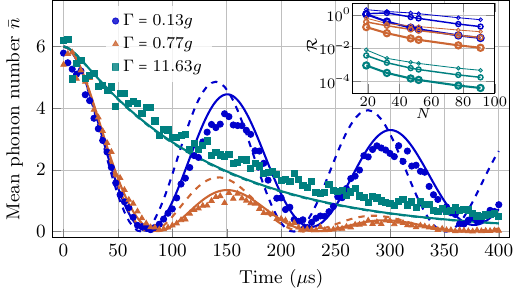}
\caption{\justifying Highest-frequency non-COM mode population dynamics of a $N=91$ planar ion crystal under collective red sideband drive in different dissipation regimes. Dashed lines represent the analytical large-$N$ limit given by Eq.\eqref{eq:nb_an}, solid lines are dynamics from the third-order cumulant expansion. (Inset) For growing crystal size, the decreasing integrated measure $\mathcal{R}$ shows how dynamics of three first motional modes approaches the same analytical limit of Eq.~\eqref{eq:nb_an}. See the text for more details.}
\label{fig:gammas}
\end{figure}
In Fig.~\ref{fig:gammas} we show measured cooling transients of the considered mode of a planar $91$-ion crystal for three values of the dissipation rate $\Gamma$ together with the approximate numerical solutions of the many-body master equation for this mode (solid lines), and the infinite-size COM asymptotics (dashed lines) of Eq.~\eqref{eq:nb_an}. The mean phonon number $\bar n$ is measured via a coherent state swap technique that will be discussed later. For $N=91$ ions and inhomogeneous spin-to-mode couplings, the direct integration of the full master equation is out of reach. To compute the dynamics, we therefore use a cumulant expansion, truncating fourth-order correlations by setting the corresponding cumulants to zero \cite{supplemental}. For the parameter regime considered here, this approach is more efficient and performs better than a discrete truncated-Wigner approximation \cite{supplemental}. The microscopic model features no free parameters. The results show that the $91$-ion dynamics of the considered out-of-phase mode is close to the infinite-size COM limit. More generally, for large crystals, the detailed distribution of the mode-dependent coupling coefficients becomes increasingly irrelevant as long as the majority of the ions participate in the dynamics. 
To illustrate this, in the inset of Fig.~\ref{fig:gammas} we consider analogous configurations of planar crystals with different sizes and plot the integrated residual $\mathcal{R}=\bar{n}_0^{-1}\int_0^{\pi/2}[\bar n^N(t)-\bar n^{\infty}(t)]^2\,\mathrm{d} gt$, where $\bar{n}^N$ are the numerical cumulant curves and $\bar{n}^{\infty}$ is given by Eq.~\eqref{eq:nb_an}. Here, the three highest-frequency motional modes are considered and the COM mode is shown by the thickest curve. In all regimes of dissipation, $\mathcal{R}$ decreases with $N$, indicating convergence toward the collective large-$N$ dynamics.

Next, we analyze finite-size effects in the collective pulsed-cooling regime starting with an effective bosonic description of the spins based on the Holstein--Primakoff picture
$J_+\simeq b^\dagger(1-b^\dagger b/2N)$. Here, we first consider the COM mode and neglect damping, $\Gamma=0$, which allows for analytical treatment. In this description, the red-sideband Hamiltonian~\eqref{eq:rsb_main} generates a coherent exchange between the motional mode and an effective collective spin mode. For an initial phonon-number distribution $\{p_n\}_{n=0}^\infty$, the time dependence of the mean phonon number is found to be
\begin{equation}
    \bar{n}(t)
    =
    \sum_{n=0}^\infty
    p_n n \cos^2(g_n t),
    \label{eq:HPnb_main}
\end{equation}
where $g_n=g(1-n/4N)$ is the effective Rabi frequency in the subspace with $n$ excitations \cite{supplemental}. For low-lying Fock states, $n\ll N$, these Rabi frequencies are nearly identical. In the limit $N\to\infty$, all effective Rabi frequencies become equal and Eq.~\eqref{eq:HPnb_main} reduces to perfect cosine oscillations, independent of the initial phonon distribution and consistent with the large-$N$ result discussed above. For finite $N$, however, the spread of Rabi frequencies prevents a perfect state swap. The residual mean occupation at $gt=\pi/2$ near the first minimum is
\begin{equation}
    \bar{n}_1
    =
    \sum_{n=0}^\infty
    p_n n
    \sin^2\left(\frac{n\pi}{8N}\right)
    =
    \frac{\pi^2}{64N^2}
    \langle n^3\rangle_0
    +
    O(N^{-4}),
    \label{eq:amin}
\end{equation}
where $\langle n^3\rangle_0=\sum_n p_n n^3$ is the third statistical moment of the initial phonon-number distribution. Hence, the cooling factor after one pulse is $\frac{\bar n_1}{\bar n_0}
    =\frac{\pi^2\langle n^2\rangle_{\rm sb}}{64N^2}    ,$
where $\langle\cdot\rangle_{\rm sb}$ denotes averaging with respect to the size-biased distribution
$q_n=n p_n/\bar n_0$. The mean phonon number is therefore reduced by a factor proportional to $1/N^2$, demonstrating a collective enhancement of pulsed sideband cooling in ion crystals. Beyond the mean occupation, further statistical properties of the motional state after the pulse can be extracted from its characteristic function, Eq.~\eqref{eq:char} \cite{supplemental}. For finite ion numbers, the residual phonon distribution is predicted to be sub-Poissonian, with $\Delta^2 n \lesssim \langle n\rangle$.

\begin{figure}[t]
\centering
\includegraphics{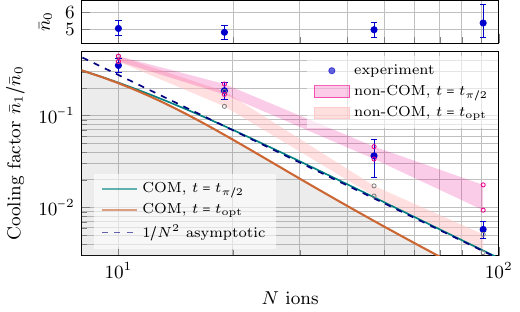}
\caption{\justifying (Upper panel) initial mode occupation $\bar{n}_0$ of the highest-frequency non-COM mode of a planar ion crystal with $N=10,19,47$ and $91$ ions. (Lower panel) cooling factor found as ratio between the initial and final mode mean occupation after one coherent collective red sideband pulse. Error bars are obtained from Monte-Carlo simulation with 95\% confidence interval. The non-COM calculations are done with cumulant expansion, the COM calculations are exact diagonalization. Dashed line is the result predicted for the COM mode with Eq.~(\ref{eq:amin}).}
\label{fig:cooling_factor}
\end{figure}

The quadratic scaling in Eq.~\eqref{eq:amin} is confirmed by experimentally measuring the cooling factor of the highest-frequency out-of-phase mode for planar crystals with $N=10,19,47$ and $91$ ions. Using the single-ion addressing capability of our setup, we can drive $|g,n\rangle\rightarrow|e,n\pm1\rangle$ sideband transitions on a selected ion. This lets us avoid self-referential verification of the cooling method by using independent techniques such as the sideband ratio method or the red-sideband excitation dynamics \cite{Leibfried2003}. After motional state preparation by sideband heating, we apply a single red-sideband pulse followed by internal state reinitialization. Due to significantly non-thermal statistics of the resulting states, the sideband ratio method cannot be used, and the mean phonon number of the cooled mode for 10, 19 and 47 ion crystals is measured via a 3rd order Taylor fit of the red sideband short-time dynamics.
For 91 ions, the final temperature is below the resolution of the fitting technique, and simultaneously the resulting state is close to the ground state and allows us to use the sideband ratio method, for which $\bar n=\frac{P_{rsb}}{P_{bsb}-P_{rsb}}$, where $P_{bsb}$ $(P_{rsb})$ are the blue and red sideband excitation probabilities. In Fig.~\ref{fig:cooling_factor}, we plot the measured cooling factor (lower panel) together with the initial mean phonon number of the mode (upper panel), prepared in a thermal state. The shaded regions indicate the expected cooling factors obtained from the cumulant theory, including the uncertainty of the initial mean phonon number; for $N=10$, exact diagonalization is used instead. For finite $N$, the time for the global minimum of Eq.~\eqref{eq:HPnb_main} has not yet converged to $t_{\pi/2}=\pi/2g$. We therefore distinguish between the cooling factor at the nominal swap time and the optimized value obtained for a $10$--$20$\% longer pulse, $t_{\rm opt}$. The measured cooling factors lie within the predicted regions for all crystal sizes and follow the approximate $1/N^2$ scaling derived above for homogeneous coupling (Eq.~(\ref{eq:amin}), dashed line). This confirms that the collective enhancement persists for the inhomogeneous modes used in the experiment. For reference, we also plot the expected cooling factor of the COM mode, obtained by solving the time dynamics exactly (solid lines).

\begin{figure}[t!]
\centering
\includegraphics{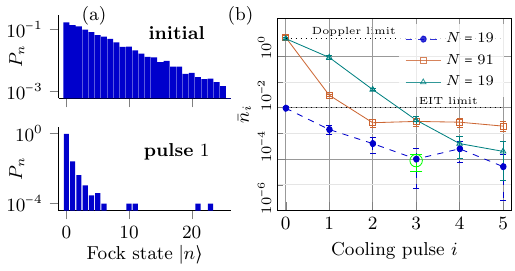}
\caption{\justifying (a) Phonon number distribution reconstructed from the measured spin counting statistics before (upper panel) and after (lower panel) one collective red sideband pulse on an inhomogeneous mode of an $91$-ion crystal. (b) Iterated pulsed cooling of the highest-frequency out-of-phase mode for planar crystals with $N=19$ and $N=91$ ions. Initial states of the mode are prepared using EIT cooling or blue sideband heating.}
\label{fig:pulsed}
\end{figure}
During the coherent pulse, the distribution of motional excitations is effectively mapped onto the internal degrees of freedom, as every motional Fock state is converted into a Dicke state \cite{hume_preparation_2009}. An adiabatic drive of the respective transition produces a similar state swap effect \cite{Lechner2016, kirkova_adiabatic_2021}, but is significantly slower than a pulse. Measuring the excited spins in the Dicke basis reconstructs the initial phonon-number distribution and provides an alternative approach for measuring the phonon mode occupation, not limited to thermal statistics. Shown in Fig. \ref{fig:pulsed}(a) are phonon distributions of the highest-frequency non-COM mode of a $91$-ion crystal, obtained using this technique before and after one cooling pulse. This method was used for temperature measurements in Fig.~\ref{fig:gammas} as the fastest available technique in the sub-Doppler domain, where the mean phonon number is much smaller than the crystal size.

The pulsed cooling can be iterated, once the excited spins are reinitialized. Fig. \ref{fig:pulsed}(b) illustrates repeated steps of cooling for two crystals, $N=19$ and $N=91$ prepared with different initial occupations using EIT cooling (dashed line) or sideband heating (solid lines). After a few cooling pulses, the mean occupation levels off due to competing processes, such as mode heating and crosstalk to the neighboring modes, with the latter limiting both cooling and measurement resolution. The observed mean phonon numbers at the level of $10^{-4}$ (the lowest stable value is $8.8^{+6.5}_{-5.3}\cdot 10^{-5}$, green circle) represent a record-low level for mechanical oscillators, to the best of our knowledge. 


To conclude, we demonstrated that collective effects are beneficial for sideband cooling of ion crystals in the regime of strong spin-motion coupling. If the crystal is large, the presented scheme for collective pulsed cooling offers enhanced cooling efficiency and reduced sensitivity to the initial phonon statistics, which applies for arbitrary motional modes and can be especially helpful when the initial motional state is non-thermal. In the latter case, the possibility to probe the mode phonon statistics and measure its effective temperature by the measurements of the spins is of particular advantage. As the length of one cooling pulse is determined solely by the tunable coupling strength, the technique can be helpful when the time budget is constrained, which is the case e.g. in ion clocks and quantum registers. By achieving exceptionally low mean phonon numbers for various ion crystals, we highlight the practical efficiency of the strong coupling regime. 

As crosstalk effects become increasingly important for larger crystals, an analysis of pulsed cooling and accurate thermometry in the multi-mode regime can be a natural extension of this approach. Another interesting direction is the possibility of collective pulsed cooling with dark states, which has not been analyzed in detail or demonstrated so far. Finally, the proposed scheme for probing the mode phonon distribution opens a new way to inspect the heating processes in ion crystals.

\begin{acknowledgments}
I.V. and K.H. acknowledge support from DFG through the Collaborative
Research Center SFB1227 (DQ-mat Project-ID 274200144) and from the Federal Ministry for Research, Technology and Space (BMFTR) Germany through project ATIQ. This project has received funding from the European Research Council (ERC) under the European Union’s Horizon 2020 research and innovation programme (grant agreement No. 101213371). Furthermore, it was funded in part by the Austrian Science Fund (FWF) [10.55776/COE1] and via the Austrian Science Fund through the Spezialforschungsbereich BeyondC (Grant No. F7110).

During the preparation of this manuscript, we became aware of a closely related theoretical study, also investigating the collective cooling mechanisms for large ion crystals \cite{shankar_new}. 
\end{acknowledgments}



%

\newpage
\appendix

\section{Experimental setup and thermometry techniques}

The experimental setup is based on a monolithic Paul trap and capable of trapping stable 2D crystals of up to 100 $^{40}$Ca$^+$ ions, with out-of-plane mode frequencies typically ranging from about 1.2 to 2.22\,MHz. Note that planar crystals even with large ion numbers around 100 require much less anisotropic potentials than linear chains with similar ion numbers. Moreover, in contrast to linear chains, the peak spectral mode density of planar crystals is not found at the upper end of the mode spectrum. Thus, we obtain comparatively large splittings of about 30\,kHz between the highest-frequency out-of-plane modes, which strongly reduces crosstalk from spectator modes in thermometry measurements. The crystal is initially Doppler-cooled followed by EIT cooling on the S$_{1/2}\leftrightarrow\mbox{P}_{1/2}$ transition of all out-of-plane modes close to the ground state \cite{kiesenhofer_controlling_2023}. A magnetic field of about 4\,Gauss lifts the degeneracy of the Zeeman sublevels. Internal  4S$_{1/2}, m_J=-1/2$ state initialization is implemented via a combination of continuous and pulsed pumping with an infidelity $<\,10^{-5}$. For coherent operations, the setup features global and addressed narrow-linewidth laser beams at 729~nm for exciting the 4S$_{1/2}\rightarrow3\mbox{D}_{5/2}$ quadrupole transition. 

Thermometry measurements are performed using three different techniques: For the data shown in Fig.~\ref{fig:gammas}, the mean phonon number is measured using the excitation mapping technique proposed in this paper, as it is the fastest, and works well in the temperature range corresponding to a few phonons. We reduce crosstalk from nearby spectator modes by carrying out the spin-motion swap slowly, employing an amplitude-shaped pulse of 1~ms duration with 100~$\mu$s ramp-up and ramp-down times. In contrast, when cooling the ions, the swap duration is 75~$\mu$s.

To avoid circular verification of the cooling method and the aforementioned errors at low temperatures, conventional thermometry methods are used for all other measurements. For thermometry close to the ground state, the sideband ratio method \cite{Leibfried2003} with an addressed 729\,nm laser is used. We select an ion that has optimal ratio between the coupling to the mode of interest vs. the closest neighboring modes in order to suppress spurious signals from the spectator modes. For the cooling efficiency measurement of 10, 19 and 47 ion crystals, the resulting state of motion after a single cooling pulse is not thermal. To measure the $\bar n$ in this case, the sideband ratio thermometry method can no longer be used, as it assumes a thermal distribution. The result would underestimate $\bar n$ in our case. To overcome this limitation, we measure red sideband excitation dynamics with an addressed laser beam and fit the data with the first three terms of the Taylor expansion $P\left(t\right)=\sum_n{P_n\left(1-\cos\left(\sqrt{n}\Omega_{RSB}t\right)\right)/2}\approx \langle n\rangle t^2\frac{\Omega_{RSB}^2}{2\cdot2!} - \langle n^2\rangle t^4\frac{\Omega_{RSB}^4}{2\cdot4!}+\langle n^3\rangle t^6\frac{\Omega_{RSB}^6}{2\cdot6!}$. Here, the first term contains the $\bar n$ and the other two are kept to improve fitting quality. The sideband Rabi frequency $\Omega_{RSB}$ is not a free parameter and is measured by inserting a single phonon into ground state cooled mode via a combination of an addressed blue sideband $\pi$-pulse and spin state reinitialization. This method yields reliable results for non-thermal distributions and is also used for measuring elevated $\bar{n} (\ge 1)$ of thermal and nonthermal distributions, where the sideband method becomes suboptimal.

\section{Anomalous heating of non-COM modes}
\begin{figure}[t]
\centering
\includegraphics{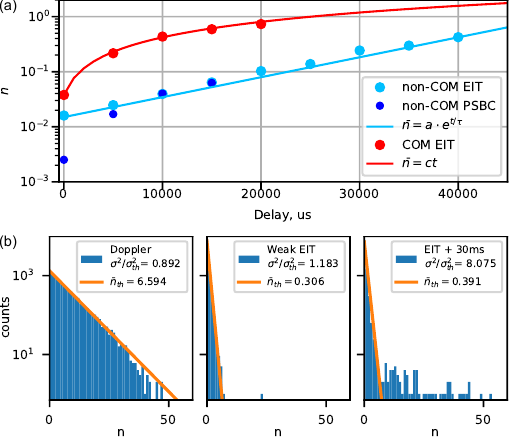}
\caption{\justifying Experimental observation of a nonlinear heating process and its statistics for the highest frequency non-COM mode in 91 ions crystal. (a): Nonlinear rise of the phonon number, $\bar{n}(t)=ae^{t/\tau}$, ($a\approx 0.015, \tau\approx 12$~ms) of $\bar{n}$ with waiting time $t$ for the non-COM mode (light-blue and blue color) and expected linear heating behavior, $\bar{n}(t)=ct$, ($c\approx 0.4$/s/ion) of the COM mode (red color). Notably, lower initial temperature obtained with pulsed sideband cooling (PSBC, blue) asymptotically gives rise to the same heating exponent. (b): The non-COM mode statistics acquired with the state mapping technique for Doppler-cooled (left), weakly EIT-cooled (middle) states, and a state observed after 30~ms of free evolution (right). The latter exhibits significantly non-thermal statistics.}
\label{fig:nonlin_heat}
\end{figure}
In our experiment we observe a rather sharp limit to the coherent operation time, after which the laser-ion interaction degrades nonlinearly. We assign this degradation to the increase in the mean occupation number $\bar{n}$ of the drumhead modes, which leads to reduced contrast of Rabi oscillations. Using the sideband ratio technique, we measure an exponential increase of $\bar{n}$ for the non-COM drumhead mode Fig. \ref{fig:nonlin_heat} (red line). In contrast, the COM mode has an expected linear increase in temperature with time and number of ions, which occurs at a rate of $~0.4$~phonons/s/ion. Using the motion-spin mapping technique proposed in this paper, we investigate the statistics of the mode states at later times. In Fig. \ref{fig:nonlin_heat} we compare phonon statistics of the non-COM mode for three cases: at the Doppler limit (left), after a purposely shortened EIT cooling stage (middle) and after complete EIT cooling followed by 30~ms of free evolution. For the first two measurements we observe statistics close to the one of a thermal state, which is indicated by the variance deviation factor $\sigma^2/\sigma^2_{th}$ being close to 1. Here $\sigma^2_{th}=\bar{n}\left(\bar{n}+1\right)$ - variance of a thermal state with the same $\bar{n}$. The variance deviation factor slightly below 1 observed for the Doppler-cooled case could be an artifact of imperfect spin-motion mapping at large phonon numbers. By contrast, the motional state found after the free evolution shows a significantly non-thermal statistics, with a strong tail at high $n$ Fock states. Such statistics could indicate a parametric excitation process. It could originate from modulation of the trap electrode potentials or parametric (nonlinear) coupling to other modes, potentially initiated by background gas collisions. Although the exact nature of this process requires further investigation, this result demonstrates the usefulness of this measurement technique; as it grants direct access to the motion state statistics, it may provide novel insights into phonon-changing processes and lead to an improvement of ion trapping technology.

\section{Effective EIT cooling model}
\label{app:eit}
Here, we describe how the balanced-beams EIT cooling is mapped onto an effective sideband cooling model and show how the regime of strong spin-motion coupling can be achieved. Similar to the analysis in Ref.~\cite{shankar}, our starting point is the Hamiltonian of $N$ three-level systems with the relevant states $\ket{e}$, $\ket{g}$ and $\ket{x}$ coupled to a well-resolved motional mode:
\begin{equation}
H_{\rm EIT}=H_{0,s}+H_{0,m}+H_1^{ex}+H_1^{eg}
\end{equation}
with
\begin{multline}
    H_{0s}=\sum_{i=1}^N\Delta_g\dyad{g_i}+\Delta_x\dyad{x_i} +\\+ \frac{\Omega_{eg}}{2}(\dyad{e_i}{g_i}+h.c.)+\frac{\Omega_{ex}}{2}(\dyad{e_i}{x_i}+h.c.),
\end{multline}
\begin{align}
    H_{0m}&=\nu a^\dagger a,\\
    H_1^{ex}&=\sum_{i=1}^N\lambda_i\frac{\eta\Omega_{ex}}{2}(a+a^\dagger)(\dyad{e_i}{x_i}+h.c.),\\
    H_1^{eg}&=-\sum_{i=1}^N\lambda_i\frac{\eta\Omega_{eg}}{2}(a+a^\dagger)(\dyad{e_i}{g_i}+h.c.).
\end{align}
Here, $\Omega_{eg},\Omega_{ex}$ and $\Delta_g, \Delta_x$ are the Rabi frequencies and the detunings of the corresponding transitions, $\nu$ is the mode frequency, $\eta$ is the Lamb-Dicke parameter, and $\{\lambda_i\}_{i=1}^N$ are the mode cooperativity factors. The incoherent part of the dynamics is governed by the two Lindblad terms corresponding to the two decay channels of the excited state:
\begin{align}
    \Gamma_{ge}\sum_{i=1}^N\mathcal{D}[\dyad{g_i}{e_i}],\quad \Gamma_{xe}\sum_{i=1}^N\mathcal{D}[\dyad{x_i}{e_i}].
\end{align}
The bare spin part $H_{0s}$ can be diagonalized and the interaction Hamiltonian can be expressed in terms of the \emph{dressed} basis $\{\ket{B_+},\ket{B_-},\ket{D}\}$ (ion index omitted), see Ref.~\cite{monroe}:
\begin{align}
    \ket{g}&=\frac{\Omega_{ex}}{\Omega}\ket{D}+\frac{\Omega_{eg}}{\sqrt{2\Omega'}}\left(\frac{\ket{B_+}}{\sqrt{\Omega'-\Delta}}-\frac{\ket{B_-}}{\sqrt{\Omega'+\Delta}}\right),\\
    \ket{x}&=-\frac{\Omega_{eg}}{\Omega}\ket{D}+\frac{\Omega_{ex}}{\sqrt{2\Omega'}}\left(\frac{\ket{B_+}}{\sqrt{\Omega'-\Delta}}-\frac{\ket{B_-}}{\sqrt{\Omega'+\Delta}}\right),\\
    \ket{e}&=\frac{\Omega}{\sqrt{2\Omega'}}\left(\frac{\ket{B_+}}{\sqrt{\Omega'+\Delta}}+\frac{\ket{B_-}}{\sqrt{\Omega'-\Delta}}\right).
\end{align}
The coupling terms are written as:
\begin{multline}
        H_1^{ex}=\frac{\eta\Omega_{ex}}{2}(a+a^\dagger)\sum_{i=1}^N\lambda_i\Big\{-\frac{\Omega_{eg}}{\sqrt{2\Omega'(\Omega'+\Delta)}}\dyad{B_{+,i}}{D_i}\\-\frac{\Omega\Omega_{ex}}{\Omega'^2-\Delta^2}\dyad{B_{+,i}}{B_{-,i}}+\frac{\Omega_{eg}}{\sqrt{2\Omega'(\Omega'-\Delta)}}\dyad{D_i}{B_{-,i}}+h.c.\Big\},\\
\end{multline}
\begin{multline}
    H_1^{eg}=-\frac{\eta\Omega_{eg}}{2}(a+a^\dagger)\sum_{i=1}^N\lambda_i\Big\{\frac{\Omega_{ex}}{\sqrt{2\Omega'(\Omega'+\Delta)}}\dyad{B_{+,i}}{D_i}\\-\frac{\Omega\Omega_{eg}}{\Omega'^2-\Delta^2}\dyad{B_{+,i}}{B_{-,i}}-\frac{\Omega_{ex}}{\sqrt{2\Omega'(\Omega'-\Delta)}}\dyad{D_i}{B_{-,i}}+h.c.\Big\}.
\end{multline}
Here, we assumed $\Delta_{eg}=\Delta_{ex}=\Delta$. This makes the dressed state energies: $E_D=\Delta$, $E_{B_{\pm}}=(\Delta\pm\Omega')/2$, where $\Omega'=\sqrt{\Delta^2+\Omega^2}$ and $\Omega=\sqrt{\Omega_{eg}^2+\Omega_{ex}^2}$. Next, we impose a condition of balanced EIT-drives: $\Omega_{eg}=\Omega_{ex}\equiv\Omega_0$ \cite{morigi}. With this, the terms coupling the states $\ket{B_+}$ and $\ket{B_-}$ cancel out in the resulting interaction Hamiltonian $H_1^{eg}+H_1^{ex}$ and it acquires the following form:
\begin{align}
    H_{1}=\eta(a+a^\dagger)\sum_{i=1}^N\lambda_i\Big\{\Omega_a\dyad{B_{+,i}}{D,i}+\Omega_b\dyad{D_i}{B_{-,i}}+h.c.\Big\},
\end{align}
where we introduced $\Omega_a=-\frac{\Omega_0^2}{\sqrt{2\Omega'(\Omega'+\Delta)}}$ and $\Omega_b=\frac{\Omega_0^2}{\sqrt{2\Omega'(\Omega'-\Delta)}}$. Next, we go to the interaction picture with respect to the bare part $H_{0m}+H_{0s}$. The full Hamiltonian thus reduces to only the interaction part and acquires time dependence. From now on, we set $\lambda_i=1/\sqrt{N}$ and for simplicity only focus on the dynamics of the COM mode of motion:
\begin{multline}
    H=\eta(ae^{-\i\nu t}+a^\dagger e^{\i\nu t})\frac{1}{\sqrt{N}}\sum_{i=1}^N\Big\{\Omega_a\dyad{B_{+,i}}{D,i}e^{\i\frac{\Omega'-\Delta}{2}t}\\+\Omega_b\dyad{D_i}{B_{-,i}}e^{\i\frac{\Omega'+\Delta}{2}t}+h.c.\Big\}.
    \label{eq:RWAham}
\end{multline}
In the above Hamiltonian \eqref{eq:RWAham}, terms of four different frequencies contribute:
\begin{align}
    \omega_1=\frac{1}{2}(\Omega'-\Delta)-\nu,\,\,
    \omega_2=\frac{1}{2}(\Omega'-\Delta)+\nu,\notag \\
    \omega_3=\frac{1}{2}(\Omega'+\Delta)-\nu,\,\,
    \omega_4=\frac{1}{2}(\Omega'+\Delta)+\nu.\notag
\end{align}
The EIT cooling condition tunes the AC Stark shift of the excited state to match the motional frequency, implying $\omega_1\approx 0$. For the remaining frequencies, we assume $\omega_2 \ll \omega_{3,4}$ and neglect the fast timescales in a rotating wave approximation, which removes the state $\ket{B_{-,i}}$ from the Hamiltonian. For the incoherent part, it can as well be shown that the contributions involving $\ket{B_{-,i}}$ can be neglected. Thus, the effective Hilbert subspace of each ion is spanned by the two states $\ket{B_{+,i}}$ and $\ket{D_i}$. The resulting Hamiltonian is a Dicke-type interaction with a resonant red sideband and an off-resonant blue sideband contribution:
\begin{align}
    H_{rb}=\frac{g}{\sqrt{N}}\sum_{i=1}^N(a\sigma_i^++h.c.)-\frac{g}{\sqrt{N}}\sum_{i=1}^N(a\sigma_i^-e^{-2\i\nu t}+h.c.),
    \label{eq:ham_dicke}
\end{align}
where we have introduced the \emph{effective} coupling parameter $g=-\eta\Omega_a=\frac{\eta\Omega_0^2}{\sqrt{2\Omega'(\Omega'+\Delta)}}$ and the spin ladder operator $\sigma_i^-=\dyad{D_i}{B_{+,i}}$. Note that there is no carrier contribution in \eqref{eq:ham_dicke}. The effective sideband cooling parameters are not straightforwardly related to those of the initial EIT system. Next, we will now show that the strong spin-motion coupling can still be achieved.

Let us calculate the incoherent term corresponding to the collapse into one of the ground states for one of the ions (index omitted) in the interaction picture $U=e^{-\i(H_{0m}+H_{0s})t}$:
\begin{align}
        \mathcal{D}\left[U\dyad{g}{e}U^\dagger\right]\approx\left(\frac{\Omega_a}{\Omega_0}\right)^2\mathcal{D}[\sigma^-]+\left(\frac{\Omega_0\Omega}{2\Omega'\sqrt{\Omega'^2-\Delta^2}}\right)^2\mathcal{D}[\sigma^{B_+}],
        \label{eq:collapses}
\end{align}
where we introduced $\sigma^{B_+}=\dyad{B_+}$. The cross-terms appearing due to the non-linearity of the dissipator are neglected, given $\Omega_0\gg\nu$. In a typical EIT cooling experiment, the bare drive Rabi frequency $\Omega_0$ has values of several dozens of MHz and the mode frequency $\nu$ is around several MHz \cite{jordan,monroe}, ensuring the validity of the rotating wave approximation. With this assumption, the action of the second jump operator $\dyad{x}{e}$ results in the same expression as in Eq.\eqref{eq:collapses}. Thus, the full incoherent dynamics is captured by the two Lindblad terms: $\Gamma_1\sum_{i=1}^N\mathcal{D}[\sigma^-_i],\quad \Gamma_{2}\sum_{i=1}^N\mathcal{D}[\sigma^{B_+}_i]$ with the corresponding rates:
\begin{equation}
    \Gamma_1=\bar{\Gamma}\left(\frac{\Omega_a}{\Omega_0}\right)^2,\quad\Gamma_2=\bar{\Gamma}\left(\frac{\Omega_0\Omega}{2\Omega'\sqrt{\Omega'^2-\Delta^2}}\right)^2
    \label{eq:eff_rates}
\end{equation}
where $\bar{\Gamma}=\Gamma_{ge}+\Gamma_{xe}$. If the EIT cooling condition is satisfied, $\Omega'=2\nu+\Delta$, the effective decay rate $\Gamma_1$ of the dressed two-level system can be written as:
\begin{align}
    \Gamma_1=\bar{\Gamma}\left(\frac{\Omega_a}{\Omega_0}\right)^2=\frac{\bar{\Gamma}}{2}\frac{1}{1+\frac{1}{2}\left(\frac{\Omega_0}{\nu}\right)^2}\approx\bar{\Gamma}\left(\frac{\nu}{\Omega_0}\right)^2.
    \label{eq:inc_rate1}
\end{align}
The rate of the effective dephasing is found as:
\begin{align}
    \Gamma_2=\bar{\Gamma}\left(\frac{\Omega_0\Omega}{2\Omega'\sqrt{\Omega'^2-\Delta^2}}\right)^2=\frac{\bar{\Gamma}}{8}\frac{1}{1+\frac{1}{2}\left(\frac{\Omega_0}{2\nu}-\frac{\nu}{\Omega_0}\right)^2}\approx\bar{\Gamma}\left(\frac{\nu}{\Omega_0}\right)^2.
    \label{eq:inc_rate2}
\end{align}
Thus, the two incoherent rates (\ref{eq:inc_rate1},\ref{eq:inc_rate2}) are related:
\begin{align}
    \Gamma_1\approx\Gamma_2\approx\bar{\Gamma}\frac{\nu^2}{\Omega_0^2}.
\end{align}
The coupling strength attains a simple form that does not depend on the initial Rabi frequency:
\begin{align}
    g=\frac{\eta\Omega_0^2}{\sqrt{2\Omega'(\Omega'+\Delta)}}=\frac{\eta\Omega_0}{2\sqrt{1+\left(\frac{\Omega_0}{2\nu}\right)^2}}\approx\eta\nu
\end{align}
Using the above expressions, we find the ratio between the effective coupling strength and the decay rate:
\begin{align}
    \frac{g}{\Gamma_1}=\frac{\eta\Omega_0}{\bar{\Gamma}}\cdot\frac{\Omega_0}{\nu}.
    \label{eq:gGratio}
\end{align}
With the typical values of $\bar{\Gamma}$ ranging from several to several dozens of MHz and the small Lamb-Dicke parameter, $\eta\ll 1$, the first multiplier in \eqref{eq:gGratio} is small: $\eta\Omega_0/\bar{\Gamma}\ll 1$. Although the second multiplier remains large, it can be possible to strike the balance between the two terms to achieve the strong coupling regime $g/\Gamma_1 \gtrsim 1$. For example, adopting the parameters used in Ref.\cite{shankar} (Fig. 7), $\Delta=360\times 2\pi$ MHz, $\Omega_{0}=33.9\times 2\pi$ MHz, $\Gamma_{eg}=6\times 2\pi$ MHz, $\Gamma_{ex}=12\times 2\pi$ MHz, $\nu=1.59\times 2\pi$ MHz, $\eta=0.066$, results in effective values $g=0.104\times 2\pi$ MHz and $\Gamma_1=0.0394\times 2\pi$ MHz and their ratio $g/\Gamma_1\approx2.64$.

\section{Dynamics in Holstein-Primakoff limit}
\label{app:hp}
For the COM mode, all spins are coupled equally to the bosonic mode: $\lambda_i=1/\sqrt{N}\,\,\forall \,i=1\dots N$. In this case, the spin-boson Hamiltonian \eqref{eq:rsb_main} takes form of a Tavis-Cummings model:
\begin{equation}
    H_{\rm TC}=\frac{g}{\sqrt{N}}(a\tilde{J}_+ + h.c.),
\end{equation}
with $\tilde{J}=\sum_{i=1}^N\sigma^+_i$. Given that the total number of excitations in the spins remains small compared to the number of ions $N$, Holstein-Primakoff transformation maps the collective spin onto an effective bosonic mode, described by creation and annihilation operators $b$ and $b^\dag$: $\tilde{J}_+\approx\sqrt{N}b^\dag(1-\frac{b^\dag b}{2N})$. For a subspace with $n$ excitations in the system, the dynamics is generated by an effective Hamiltonian:
\begin{equation}
    H_n=g_n(ba^\dag+h.c.),
    \label{eq:indiv}
\end{equation}
where $g_n=g(1-\frac{b^\dag b}{4N})$ is the individual Rabi frequency. Hamiltonian \eqref{eq:indiv} describes a beam-splitter interaction between the two bosonic modes. To analyze it further, we introduce a Schwinger spin notation with the following operators:
\begin{align}
    &S_+=ba^\dag,\\
    &S_z=\frac{1}{2}(a^\dag a - b^\dag b)\\
    &S_0=\frac{1}{2}(a^\dag a + b^\dag b).
\end{align}
The initial system state for a subspace with $n$ excitations corresponds to the polarized state of the Schwinger spin: $\dyad{n}\otimes\dyad{\frac{N}{2},-\frac{N}{2}}\equiv\ket{\frac{n}{2},\frac{n}{2}}_S$. The mean phonon number is found as a weighted sum of the Schwinger spin rotations about the $x$-axis:
\begin{equation}
    \langle a^\dag a \rangle = \sum_{n=0}^{\infty}p_n \Tr_{\text{sm}}\left[ (S_0+S_z)U(t)\ket{\frac{n}{2},\frac{n}{2}}_S\bra{\frac{n}{2},\frac{n}{2}}U^\dag(t) \right]
\end{equation}
with $U(t)=e^{-ig_n t(S_++S_-)}$ and $\{p_n\}_{n=0}^\infty$ being the initial phonon distribution. A further straightforward calculation yields an analytical result:
\begin{equation}
    \langle a^\dag a \rangle =\frac{1}{2}\left( \bar{n}_0+\sum_{n=0}^\infty p_n n \cos(2g_n t)\right)=\sum_{n=0}^\infty p_n n\cos^2\left(g_nt\right),
    \label{eq:HPnbar}
\end{equation}
where $\bar{n}_0$ is the initial mean phonon number. For a cold and large ion crystal, the phonon distribution is concentrated in the low-lying Fock states, $\exists\, n^*\ll N: p_{n>n^*}\approx 0$, and the relevant Schwinger rotations synchronize: $g_{n<n^*}\approx g$. In this limit, the results are perfect cosine oscillations of the mode mean occupation number:
\begin{equation}
    \langle a^\dag a \rangle = \bar{n}_0\cos^2(gt).
\end{equation}
At $gt=\pi/2$ the two coupled oscillators have not only exchanged energy, but also their quantum states. This allows one to probe the properties of the motional mode $a$ via the accessible spin mode $b$ by measuring the latter in the Dicke basis.

\section{Resulting motional state}
\label{app:mot_state}
In the Holstein-Primakoff approximation, the Fock state probability distribution $\{p'_m\}_{m=0}^{\infty}$ after time $t$ can be found as:
\begin{equation}
    p'_m=\sum_{n\geq m}p_n |z|^2,
\end{equation}
where
\begin{equation}
    z=\bra{\frac{n}{2},-\frac{n}{2}+m}e^{-\i g_n t S_x}\ket{\frac{n}{2},\frac{n}{2}}
\end{equation}
and $\{p_n\}_{n=0}^{\infty}$ is the initial Fock state distribution. To gain more insights on the properties of the state, consider the characteristic function of the output Fock state distribution:
\begin{multline}
    \chi(\lambda)=\sum_{m=0}^\infty e^{\i m\lambda}p'_m=\sum_{n=0}^{\infty} p_n e^{\i\lambda/2} \mathcal{U}^{\frac{n}{2}}_{\frac{n}{2},\frac{n}{2}}\left(-\lambda;g_nt,\frac{\pi}{2}\right),
\end{multline}
where the functions $\mathcal{U}^J_{M,M'}(\omega;\Theta,\Phi)$ are the matrix elements of a spin rotation operator (see, for example, \cite{varshalovich}). The explicit form of the characteristic function is:
\begin{multline}
    \chi(\lambda)=\sum_{n=0}^\infty p_n \exp\left\{\i n\left(\frac{\lambda}{2}+\arctan(\tan\frac{\lambda}{2}\cos 2g_n t)\right)\right\}\times\\
    \times(1-\sin^2\frac{\lambda}{2}\sin^22g_n t)^{\frac{n}{2}}
    \label{eq:char}
\end{multline}
In the large-$N$ limit, the characteristic function at $gt=\pi/2$ is $\chi(\lambda)=1$, which after taking the inverse Fourier transform corresponds to the perfect motional ground state: $p'_m=\delta_{m,0}$. Each further even-power $1/N$ correction modifies this result by adding the next non-zero Fock state contribution that depends on the values of higher-order moments of the initial distribution. For example, expanding Eq.\eqref{eq:char} to the leading order $1/N^2$ gives a probability distribution with only the two lowest Fock states populated:
\begin{equation}
    p'_m=\delta_{m,0}+\frac{\pi^2}{64N^2}\langle n^3\rangle_0\left(\delta_{m,1}-\delta_{m,0}\right),
\end{equation}
consistent with the result of Eq.\eqref{eq:amin}. Using derivatives of the characteristic function \eqref{eq:char}, one can extract all statistical moments of the resulting distribution. The mean and the variance acquire a particularly simple form:
\begin{align}
    \langle n\rangle &= \sum_{n=0}^\infty p_n n \sin^2\left(\frac{n \pi}{8 N}\right),\\
    \Delta^2n &= \frac{1}{4}\sum_{n=0}^\infty p_n n \sin^2\left(\frac{n\pi}{4 N}\right).
\end{align}
One sees that generally $\langle n\rangle>\Delta^2n$, implying a sub-Poissonian statistics for the output state. However, for a large crystal, keeping only the leading order $1/N^2$ results in a Poissonian statistics: $\langle n\rangle\approx\Delta^2n$, resembling the result for a thermal state with the same energy: $\Delta^2n_{\text{th}}=\langle n\rangle^2+\langle n\rangle\approx\langle n\rangle$.

\section{Cumulant Expansion}
\label{app:cumulants}
Cumulant expansion (also referred to as higher order mean field theory, cluster expansion or the Bogoliubov-Born-Green-Kirkwood-Yvon hierarchy) provides a way to extract the dynamics of many-body systems by systematically truncating the statistical correlations between the quantum observables; this is achieved by setting the respective statistical cumulants to zero. 
The starting point are the Heisenberg equations of motion for a relevant observable $\mathcal{O}$:
\begin{align}
    \partial_t \langle \mathcal{O}\rangle=-\i \langle[\mathcal{O},H]\rangle + \langle \mathcal{D}^\dagger[L](\mathcal{O})\rangle,
    \label{eq:cumulant_example}
\end{align}
where we assumed the dynamics of a Lindblad master equation with Hamiltonian $H$ and a collapse operator $L$. In any non-trivial case, the right hand side of Eq.\eqref{eq:cumulant_example} couples inevitably to means of higher-order operator products. To close the resulting system of equations, one expresses the higher order moments in terms of the lower order means by setting the respective cumulants to zero at a chosen order. In our particular case of the Dicke-type interaction \eqref{eq:rsb_main}, increasing the cumulant truncation order improves the approximation. However, this might not always be the case and general criteria for applicability and convergence of cumulant expansions are yet to be understood \cite{kerber_cumulants_2025}.

Apart from the purely numerical results, the obtained system might also provide analytical insights into the dynamics. For master equation \eqref{eq:meq} and the symmetric COM mode, the second-order truncated system reads (see e.g. \cite{kirton_introduction_2019} for the general case):
\begin{align}
    \partial_t\bar{n}&=-2g\sqrt{N} \operatorname{Im}\langle a\sigma^+\rangle,\\
    \partial_t\operatorname{Im}\langle a\sigma^+\rangle&=-\frac{g}{\sqrt{N}}((N-1)\langle\sigma^+\sigma^-\rangle+\\
    &+\bar{n}(2\langle\sigma^{ee}\rangle-1)+\langle\sigma^{ee}\rangle)-\frac{1}{2}\Gamma\operatorname{Im}\langle a\sigma^+\rangle,\notag\\
    \partial_t\langle\sigma^{ee}\rangle&=2\frac{g}{\sqrt{N}}\operatorname{Im}\langle a\sigma^+\rangle-\Gamma\langle \sigma^{ee}\rangle,\\
    \partial_t\langle\sigma^+\sigma^-\rangle&=2\frac{g}{\sqrt{N}}\left(1-2\langle\sigma^{ee}\rangle \right)\operatorname{Im}\langle a\sigma^+\rangle-\Gamma\langle \sigma^+\sigma^-\rangle.
\end{align} 
By rescaling the variables, $\operatorname{Im}\langle a\sigma^+\rangle\rightarrow\sqrt{N}\operatorname{Im}\langle a\sigma^+\rangle, \, \langle\sigma^{ee}\rangle\rightarrow N \langle\sigma^{ee}\rangle$ and $\langle\sigma^+\sigma^-\rangle \rightarrow N\langle\sigma^+\sigma^-\rangle$, the above system of equations becomes linear in the large-$N$ limit. Organizing the new variables in a vector $\vec{w}$, results in a matrix equation:
\begin{equation}
    \dot{\vec{w}}=M\vec{w},
\end{equation}
with
\begin{equation}
    M=\begin{pmatrix}
    0&-2g&0&0\\
    g&-\frac{\Gamma}{2}&-g&0\\
    0&2g&-\Gamma &0\\
    0&2g&0&-\Gamma
    \end{pmatrix}.
\end{equation}
If all the spins are prepared in the ground state and the mode on average has $\bar{n}_0$ phonons, the initial state of the system reads $\vec{w}_0=(\bar{n}_0,\,0,\,0,\,0)^T$, and the mode population dynamics is found analytically by matrix exponentiation:
\begin{equation}
    \bar{n} (\tau)=\frac{\bar{n}_0e^{-\tau/2}}{2\omega^2}\left(\omega^2-1+(1+\omega^2)\cosh\frac{\omega\tau}{2}+2\omega\sinh\frac{\omega\tau}{2}\right),
    \label{eq:app_transient}
\end{equation}
where $\omega=\sqrt{1-16(\frac{g}{\Gamma})^2}$ and $\tau=\Gamma t$. Note that the initial phonon number only enters as a global prefactor. For weak coupling, $g\ll\Gamma$, the time dynamics of Eq.\eqref{eq:app_transient} is dominated by an exponential decay with the rate $\frac{\Gamma}{2}(\omega-1)\approx-4\frac{g^2}{\Gamma}$, which is a sideband cooling rate of a single ion.

In the similar spirit, the second-order theory can be used to find the lower temperature limit set by an off-resonant coherent excitation of the blue sideband. For this, we consider the two-sideband COM Hamiltonian \eqref{eq:ham_dicke} with $g/\nu\ll1$ and write down the linearized cumulant equations in a matrix form:
\begin{equation}
    \dot{\vec{w}} = (R_r+R_b)\vec{w}+\vec{f_b},
    \label{eq:matrix_rsb_bsb}
\end{equation}
where $R_r$ and $R_b$ are matrices containing the contributions of the red and blue sidebands respectively and $\vec{f_b}$ is the arising inhomogeneity vector. For $\Gamma=0$, the system \eqref{eq:matrix_rsb_bsb} can be solved analytically: $\vec{w}(t)=e^{(R_r+R_b)t}(\vec{w}_0+\vec{f_b})$. We identify that in the large-$N$ limit it is the inhomogeneous part of the solution that is responsible for the fundamental cooling limit independent on the initial phonon occupation. At $gt=\pi/2$, this contribution takes form $\bar{n}_\text{lim}\approx\frac{k^2}{2}\left(1+\cos{\frac{\pi}{k}}\right)$ with $k=g/\nu$. Dropping the oscillating term results in a limit set by the off-resonant blue sideband excitation:
\begin{equation}
    \bar{n}_\text{lim}\sim\left(\frac{g}{\nu}\right)^2.
\end{equation}

For inhomogeneous modes, the variables related to different spins are independent, and the systems of cumulant equations become large. For instance, inhomogeneous dynamics of $N=91$ ions in Fig.\ref{fig:gammas} was computed using third-order truncation theory and required about $700$ thousand variables to be tracked. To derive and solve cumulant systems this large, we developed a parallelized automated procedure based on the cumulant generating function and utilized the high-performance computing capabilities offered by the University of Hannover. The exact technical details of this technique will be a part of a forthcoming publication.

\section{Discrete Truncated Wigner Approximation}
\label{app:dtwa}

\begin{figure}[t!]
\centering
\includegraphics{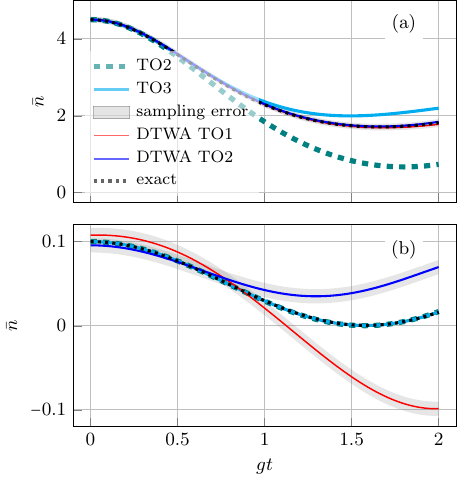}
\caption{\justifying Collective red sideband dynamics of Hamiltonian \eqref{eq:rsb_main} for an out-of-phase mode of an $N=10$ ion crystal. (a) The dynamics generated by the discrete truncated Wigner method agrees well with the exact solution in the regime of high initial phonon occupation. (b) For low initial phonon occupations near the ground state, the cumulant expansion offers a much more stable approximation.}
\label{fig:DTWA}
\end{figure}

To compute the coherent dynamics of the inhomogeneous modes, we alternatively try a sampling-based many-body numerical method known as Discrete Truncated Wigner Approximation (DTWA) \cite{pineiro_orioli_nonequilibrium_2017,schachenmayer_many-body_2015}. At the heart of the approach is the Wigner-Weyl transformation, mapping the quantum operators onto their corresponding phase space functions. For bosons, this is achieved by a standard continuous transformation and for spins the discrete formalism of Wooters is used \cite{wootters_wigner-function_1987}. The time evolution of the relevant Weyl symbols is then found using mean field theory or the next-order cumulant expansions. Each computed trajectory corresponds to a set of initial conditions, which are sampled according to the Wigner (quasi-)probability distribution, given the latter is positive over its domain. We can easily compare DTWA with the usual cumulant expansion approach (see the corresponding section), since the resulting equations of motion have the same structure. A similar comparison was made in Ref.~\cite{pineiro_orioli_nonequilibrium_2017} for a different type of spin-boson interaction, and DTWA was shown to deliver a qualitatively much better result.

In Fig.~\ref{fig:DTWA}, we show the mode population dynamics governed by the spin-boson Hamiltonian \eqref{eq:rsb_main} for the highest-frequency non-COM drumhead mode of a planar $N=10$ ion crystal. The dynamics are computed for high ($\bar{n}_0=4.5$) and low ($\bar{n}_0=0.1$) initial phonon occupations using the cumulant expansion of truncation orders (TO) $2$ and $3$ and DTWA of orders $1$ (mean-field) and $2$ with $\mathcal{N}_s=10^4$ samples. We compare these results with the exact numerical solution.

Far from the ground state, already the mean-field DTWA reproduces the exact mode population dynamics well, demonstrating an overall better agreement than the cumulant expansion. However, near the ground state the situation is opposite: the cumulant expansions of orders $2$ and $3$ overlap perfectly with the exact curve, while DTWA struggles to reproduce it correctly. First-order DTWA produces non-physical negative temperature values, which does not occur for the cumulant expansions. Increasing the truncation order consistently improves the DTWA agreement, but also tremendously increases the computational demands. In case of cumulants, no sampling is needed, so the system of equations only needs to be solved once. This motivated us to employ cumulant expansion as a more reliable and resource-friendly method for our system throughout this study. Furthermore, incorporating incoherent effects is done straightforwardly with the cumulant equations, whereas for the DTWA, modifications of the technique need to be used \cite{huber_realistic_2022,hosseinabadi_user_friendly_2025}.


\begin{thebibliography}{54}%
\makeatletter
\providecommand \@ifxundefined [1]{%
 \@ifx{#1\undefined}
}%
\providecommand \@ifnum [1]{%
 \ifnum #1\expandafter \@firstoftwo
 \else \expandafter \@secondoftwo
 \fi
}%
\providecommand \@ifx [1]{%
 \ifx #1\expandafter \@firstoftwo
 \else \expandafter \@secondoftwo
 \fi
}%
\providecommand \natexlab [1]{#1}%
\providecommand \enquote  [1]{``#1''}%
\providecommand \bibnamefont  [1]{#1}%
\providecommand \bibfnamefont [1]{#1}%
\providecommand \citenamefont [1]{#1}%
\providecommand \href@noop [0]{\@secondoftwo}%
\providecommand \href [0]{\begingroup \@sanitize@url \@href}%
\providecommand \@href[1]{\@@startlink{#1}\@@href}%
\providecommand \@@href[1]{\endgroup#1\@@endlink}%
\providecommand \@sanitize@url [0]{\catcode `\\12\catcode `\$12\catcode `\&12\catcode `\#12\catcode `\^12\catcode `\_12\catcode `\%12\relax}%
\providecommand \@@startlink[1]{}%
\providecommand \@@endlink[0]{}%
\providecommand \url  [0]{\begingroup\@sanitize@url \@url }%
\providecommand \@url [1]{\endgroup\@href {#1}{\urlprefix }}%
\providecommand \urlprefix  [0]{URL }%
\providecommand \Eprint [0]{\href }%
\providecommand \doibase [0]{https://doi.org/}%
\providecommand \selectlanguage [0]{\@gobble}%
\providecommand \bibinfo  [0]{\@secondoftwo}%
\providecommand \bibfield  [0]{\@secondoftwo}%
\providecommand \translation [1]{[#1]}%
\providecommand \BibitemOpen [0]{}%
\providecommand \bibitemStop [0]{}%
\providecommand \bibitemNoStop [0]{.\EOS\space}%
\providecommand \EOS [0]{\spacefactor3000\relax}%
\providecommand \BibitemShut  [1]{\csname bibitem#1\endcsname}%
\let\auto@bib@innerbib\@empty
\bibitem [{\citenamefont {Srinivas}\ \emph {et~al.}(2021)\citenamefont {Srinivas}, \citenamefont {Burd}, \citenamefont {Knaack}, \citenamefont {Sutherland}, \citenamefont {Kwiatkowski}, \citenamefont {Glancy}, \citenamefont {Knill}, \citenamefont {Wineland}, \citenamefont {Leibfried}, \citenamefont {Wilson}, \citenamefont {Allcock},\ and\ \citenamefont {Slichter}}]{srinivas_high_fidelity_2021}%
  \BibitemOpen
  \bibfield  {author} {\bibinfo {author} {\bibfnamefont {R.}~\bibnamefont {Srinivas}}, \bibinfo {author} {\bibfnamefont {S.~C.}\ \bibnamefont {Burd}}, \bibinfo {author} {\bibfnamefont {H.~M.}\ \bibnamefont {Knaack}}, \bibinfo {author} {\bibfnamefont {R.~T.}\ \bibnamefont {Sutherland}}, \bibinfo {author} {\bibfnamefont {A.}~\bibnamefont {Kwiatkowski}}, \bibinfo {author} {\bibfnamefont {S.}~\bibnamefont {Glancy}}, \bibinfo {author} {\bibfnamefont {E.}~\bibnamefont {Knill}}, \bibinfo {author} {\bibfnamefont {D.~J.}\ \bibnamefont {Wineland}}, \bibinfo {author} {\bibfnamefont {D.}~\bibnamefont {Leibfried}}, \bibinfo {author} {\bibfnamefont {A.~C.}\ \bibnamefont {Wilson}}, \bibinfo {author} {\bibfnamefont {D.~T.~C.}\ \bibnamefont {Allcock}},\ and\ \bibinfo {author} {\bibfnamefont {D.~H.}\ \bibnamefont {Slichter}},\ }\href {https://doi.org/10.1038/s41586-021-03809-4} {\bibfield  {journal} {\bibinfo  {journal} {Nature}\ }\textbf {\bibinfo {volume} {597}},\ \bibinfo {pages} {209} (\bibinfo {year} {2021})}\BibitemShut
  {NoStop}%
\bibitem [{\citenamefont {Zarantonello}\ \emph {et~al.}(2019)\citenamefont {Zarantonello}, \citenamefont {Hahn}, \citenamefont {Morgner}, \citenamefont {Schulte}, \citenamefont {Bautista-Salvador}, \citenamefont {Werner}, \citenamefont {Hammerer},\ and\ \citenamefont {Ospelkaus}}]{zarantonello_robust_2019}%
  \BibitemOpen
  \bibfield  {author} {\bibinfo {author} {\bibfnamefont {G.}~\bibnamefont {Zarantonello}}, \bibinfo {author} {\bibfnamefont {H.}~\bibnamefont {Hahn}}, \bibinfo {author} {\bibfnamefont {J.}~\bibnamefont {Morgner}}, \bibinfo {author} {\bibfnamefont {M.}~\bibnamefont {Schulte}}, \bibinfo {author} {\bibfnamefont {A.}~\bibnamefont {Bautista-Salvador}}, \bibinfo {author} {\bibfnamefont {R.}~\bibnamefont {Werner}}, \bibinfo {author} {\bibfnamefont {K.}~\bibnamefont {Hammerer}},\ and\ \bibinfo {author} {\bibfnamefont {C.}~\bibnamefont {Ospelkaus}},\ }\href {https://doi.org/10.1103/PhysRevLett.123.260503} {\bibfield  {journal} {\bibinfo  {journal} {Physical Review Letters}\ }\textbf {\bibinfo {volume} {123}},\ \bibinfo {pages} {260503} (\bibinfo {year} {2019})}\BibitemShut {NoStop}%
\bibitem [{\citenamefont {Löschnauer}\ \emph {et~al.}(2025)\citenamefont {Löschnauer}, \citenamefont {Mosca~Toba}, \citenamefont {Hughes}, \citenamefont {King}, \citenamefont {Weber}, \citenamefont {Srinivas}, \citenamefont {Matt}, \citenamefont {Nourshargh}, \citenamefont {Allcock}, \citenamefont {Ballance}, \citenamefont {Matthiesen}, \citenamefont {Malinowski},\ and\ \citenamefont {Harty}}]{loschnauer_scalable_2025}%
  \BibitemOpen
  \bibfield  {author} {\bibinfo {author} {\bibfnamefont {C.}~\bibnamefont {Löschnauer}}, \bibinfo {author} {\bibfnamefont {J.}~\bibnamefont {Mosca~Toba}}, \bibinfo {author} {\bibfnamefont {A.}~\bibnamefont {Hughes}}, \bibinfo {author} {\bibfnamefont {S.}~\bibnamefont {King}}, \bibinfo {author} {\bibfnamefont {M.}~\bibnamefont {Weber}}, \bibinfo {author} {\bibfnamefont {R.}~\bibnamefont {Srinivas}}, \bibinfo {author} {\bibfnamefont {R.}~\bibnamefont {Matt}}, \bibinfo {author} {\bibfnamefont {R.}~\bibnamefont {Nourshargh}}, \bibinfo {author} {\bibfnamefont {D.}~\bibnamefont {Allcock}}, \bibinfo {author} {\bibfnamefont {C.}~\bibnamefont {Ballance}}, \bibinfo {author} {\bibfnamefont {C.}~\bibnamefont {Matthiesen}}, \bibinfo {author} {\bibfnamefont {M.}~\bibnamefont {Malinowski}},\ and\ \bibinfo {author} {\bibfnamefont {T.}~\bibnamefont {Harty}},\ }\href {https://doi.org/10.1103/h4wk-v31j} {\bibfield  {journal} {\bibinfo  {journal} {PRX Quantum}\ }\textbf {\bibinfo {volume} {6}},\ \bibinfo {pages} {040313}
  (\bibinfo {year} {2025})}\BibitemShut {NoStop}%
\bibitem [{\citenamefont {Guo}\ \emph {et~al.}(2024)\citenamefont {Guo}, \citenamefont {Wu}, \citenamefont {Ye}, \citenamefont {Zhang}, \citenamefont {Lian}, \citenamefont {Yao}, \citenamefont {Wang}, \citenamefont {Yan}, \citenamefont {Yi}, \citenamefont {Xu}, \citenamefont {Li}, \citenamefont {Hou}, \citenamefont {Xu}, \citenamefont {Guo}, \citenamefont {Zhang}, \citenamefont {Qi}, \citenamefont {Zhou}, \citenamefont {He},\ and\ \citenamefont {Duan}}]{guo_site_resolved_2024}%
  \BibitemOpen
  \bibfield  {author} {\bibinfo {author} {\bibfnamefont {S.-A.}\ \bibnamefont {Guo}}, \bibinfo {author} {\bibfnamefont {Y.-K.}\ \bibnamefont {Wu}}, \bibinfo {author} {\bibfnamefont {J.}~\bibnamefont {Ye}}, \bibinfo {author} {\bibfnamefont {L.}~\bibnamefont {Zhang}}, \bibinfo {author} {\bibfnamefont {W.-Q.}\ \bibnamefont {Lian}}, \bibinfo {author} {\bibfnamefont {R.}~\bibnamefont {Yao}}, \bibinfo {author} {\bibfnamefont {Y.}~\bibnamefont {Wang}}, \bibinfo {author} {\bibfnamefont {R.-Y.}\ \bibnamefont {Yan}}, \bibinfo {author} {\bibfnamefont {Y.-J.}\ \bibnamefont {Yi}}, \bibinfo {author} {\bibfnamefont {Y.-L.}\ \bibnamefont {Xu}}, \bibinfo {author} {\bibfnamefont {B.-W.}\ \bibnamefont {Li}}, \bibinfo {author} {\bibfnamefont {Y.-H.}\ \bibnamefont {Hou}}, \bibinfo {author} {\bibfnamefont {Y.-Z.}\ \bibnamefont {Xu}}, \bibinfo {author} {\bibfnamefont {W.-X.}\ \bibnamefont {Guo}}, \bibinfo {author} {\bibfnamefont {C.}~\bibnamefont {Zhang}}, \bibinfo {author} {\bibfnamefont {B.-X.}\ \bibnamefont {Qi}}, \bibinfo
  {author} {\bibfnamefont {Z.-C.}\ \bibnamefont {Zhou}}, \bibinfo {author} {\bibfnamefont {L.}~\bibnamefont {He}},\ and\ \bibinfo {author} {\bibfnamefont {L.-M.}\ \bibnamefont {Duan}},\ }\href {https://doi.org/10.1038/s41586-024-07459-0} {\bibfield  {journal} {\bibinfo  {journal} {Nature}\ }\textbf {\bibinfo {volume} {630}},\ \bibinfo {pages} {613} (\bibinfo {year} {2024})}\BibitemShut {NoStop}%
\bibitem [{\citenamefont {Monroe}\ \emph {et~al.}(2021)\citenamefont {Monroe}, \citenamefont {Campbell}, \citenamefont {Duan}, \citenamefont {Gong}, \citenamefont {Gorshkov}, \citenamefont {Hess}, \citenamefont {Islam}, \citenamefont {Kim}, \citenamefont {Linke}, \citenamefont {Pagano}, \citenamefont {Richerme}, \citenamefont {Senko},\ and\ \citenamefont {Yao}}]{monroe_programmable_2021}%
  \BibitemOpen
  \bibfield  {author} {\bibinfo {author} {\bibfnamefont {C.}~\bibnamefont {Monroe}}, \bibinfo {author} {\bibfnamefont {W.}~\bibnamefont {Campbell}}, \bibinfo {author} {\bibfnamefont {L.-M.}\ \bibnamefont {Duan}}, \bibinfo {author} {\bibfnamefont {Z.-X.}\ \bibnamefont {Gong}}, \bibinfo {author} {\bibfnamefont {A.}~\bibnamefont {Gorshkov}}, \bibinfo {author} {\bibfnamefont {P.}~\bibnamefont {Hess}}, \bibinfo {author} {\bibfnamefont {R.}~\bibnamefont {Islam}}, \bibinfo {author} {\bibfnamefont {K.}~\bibnamefont {Kim}}, \bibinfo {author} {\bibfnamefont {N.}~\bibnamefont {Linke}}, \bibinfo {author} {\bibfnamefont {G.}~\bibnamefont {Pagano}}, \bibinfo {author} {\bibfnamefont {P.}~\bibnamefont {Richerme}}, \bibinfo {author} {\bibfnamefont {C.}~\bibnamefont {Senko}},\ and\ \bibinfo {author} {\bibfnamefont {N.}~\bibnamefont {Yao}},\ }\href {https://doi.org/10.1103/RevModPhys.93.025001} {\bibfield  {journal} {\bibinfo  {journal} {Reviews of Modern Physics}\ }\textbf {\bibinfo {volume} {93}},\ \bibinfo {pages} {025001}
  (\bibinfo {year} {2021})}\BibitemShut {NoStop}%
\bibitem [{\citenamefont {Qiao}\ \emph {et~al.}(2024)\citenamefont {Qiao}, \citenamefont {Cai}, \citenamefont {Wang}, \citenamefont {Du}, \citenamefont {Jin}, \citenamefont {Chen}, \citenamefont {Wang}, \citenamefont {Luan}, \citenamefont {Gao}, \citenamefont {Sun}, \citenamefont {Tian}, \citenamefont {Zhang},\ and\ \citenamefont {Kim}}]{qiao_tunable_2024}%
  \BibitemOpen
  \bibfield  {author} {\bibinfo {author} {\bibfnamefont {M.}~\bibnamefont {Qiao}}, \bibinfo {author} {\bibfnamefont {Z.}~\bibnamefont {Cai}}, \bibinfo {author} {\bibfnamefont {Y.}~\bibnamefont {Wang}}, \bibinfo {author} {\bibfnamefont {B.}~\bibnamefont {Du}}, \bibinfo {author} {\bibfnamefont {N.}~\bibnamefont {Jin}}, \bibinfo {author} {\bibfnamefont {W.}~\bibnamefont {Chen}}, \bibinfo {author} {\bibfnamefont {P.}~\bibnamefont {Wang}}, \bibinfo {author} {\bibfnamefont {C.}~\bibnamefont {Luan}}, \bibinfo {author} {\bibfnamefont {E.}~\bibnamefont {Gao}}, \bibinfo {author} {\bibfnamefont {X.}~\bibnamefont {Sun}}, \bibinfo {author} {\bibfnamefont {H.}~\bibnamefont {Tian}}, \bibinfo {author} {\bibfnamefont {J.}~\bibnamefont {Zhang}},\ and\ \bibinfo {author} {\bibfnamefont {K.}~\bibnamefont {Kim}},\ }\href {https://doi.org/10.1038/s41567-023-02378-9} {\bibfield  {journal} {\bibinfo  {journal} {Nature Physics}\ }\textbf {\bibinfo {volume} {20}},\ \bibinfo {pages} {623} (\bibinfo {year} {2024})}\BibitemShut {NoStop}%
\bibitem [{\citenamefont {Joshi}\ \emph {et~al.}(2022)\citenamefont {Joshi}, \citenamefont {Kranzl}, \citenamefont {Schuckert}, \citenamefont {Lovas}, \citenamefont {Maier}, \citenamefont {Blatt}, \citenamefont {Knap},\ and\ \citenamefont {Roos}}]{joshi_observing_2022}%
  \BibitemOpen
  \bibfield  {author} {\bibinfo {author} {\bibfnamefont {M.~K.}\ \bibnamefont {Joshi}}, \bibinfo {author} {\bibfnamefont {F.}~\bibnamefont {Kranzl}}, \bibinfo {author} {\bibfnamefont {A.}~\bibnamefont {Schuckert}}, \bibinfo {author} {\bibfnamefont {I.}~\bibnamefont {Lovas}}, \bibinfo {author} {\bibfnamefont {C.}~\bibnamefont {Maier}}, \bibinfo {author} {\bibfnamefont {R.}~\bibnamefont {Blatt}}, \bibinfo {author} {\bibfnamefont {M.}~\bibnamefont {Knap}},\ and\ \bibinfo {author} {\bibfnamefont {C.~F.}\ \bibnamefont {Roos}},\ }\href {https://doi.org/10.1126/science.abk2400} {\bibfield  {journal} {\bibinfo  {journal} {Science}\ }\textbf {\bibinfo {volume} {376}},\ \bibinfo {pages} {720} (\bibinfo {year} {2022})}\BibitemShut {NoStop}%
\bibitem [{\citenamefont {Franke}\ \emph {et~al.}(2023)\citenamefont {Franke}, \citenamefont {Muleady}, \citenamefont {Kaubruegger}, \citenamefont {Kranzl}, \citenamefont {Blatt}, \citenamefont {Rey}, \citenamefont {Joshi},\ and\ \citenamefont {Roos}}]{franke_quantum_enhanced_2023}%
  \BibitemOpen
  \bibfield  {author} {\bibinfo {author} {\bibfnamefont {J.}~\bibnamefont {Franke}}, \bibinfo {author} {\bibfnamefont {S.~R.}\ \bibnamefont {Muleady}}, \bibinfo {author} {\bibfnamefont {R.}~\bibnamefont {Kaubruegger}}, \bibinfo {author} {\bibfnamefont {F.}~\bibnamefont {Kranzl}}, \bibinfo {author} {\bibfnamefont {R.}~\bibnamefont {Blatt}}, \bibinfo {author} {\bibfnamefont {A.~M.}\ \bibnamefont {Rey}}, \bibinfo {author} {\bibfnamefont {M.~K.}\ \bibnamefont {Joshi}},\ and\ \bibinfo {author} {\bibfnamefont {C.~F.}\ \bibnamefont {Roos}},\ }\href {https://doi.org/10.1038/s41586-023-06472-z} {\bibfield  {journal} {\bibinfo  {journal} {Nature}\ }\textbf {\bibinfo {volume} {621}},\ \bibinfo {pages} {740} (\bibinfo {year} {2023})}\BibitemShut {NoStop}%
\bibitem [{\citenamefont {Filzinger}\ \emph {et~al.}(2026)\citenamefont {Filzinger}, \citenamefont {Steinel}, \citenamefont {Jiang}, \citenamefont {Bennett}, \citenamefont {Mehlstäubler}, \citenamefont {Peik},\ and\ \citenamefont {Huntemann}}]{filzinger_multi_ion_2026}%
  \BibitemOpen
  \bibfield  {author} {\bibinfo {author} {\bibfnamefont {M.}~\bibnamefont {Filzinger}}, \bibinfo {author} {\bibfnamefont {M.~R.}\ \bibnamefont {Steinel}}, \bibinfo {author} {\bibfnamefont {J.}~\bibnamefont {Jiang}}, \bibinfo {author} {\bibfnamefont {D.}~\bibnamefont {Bennett}}, \bibinfo {author} {\bibfnamefont {T.~E.}\ \bibnamefont {Mehlstäubler}}, \bibinfo {author} {\bibfnamefont {E.}~\bibnamefont {Peik}},\ and\ \bibinfo {author} {\bibfnamefont {N.}~\bibnamefont {Huntemann}},\ }\bibfield  {journal} {\bibinfo  {journal} {arXiv preprint}\ }\href {https://doi.org/10.48550/arXiv.2603.23446} {10.48550/arXiv.2603.23446} (\bibinfo {year} {2026})\BibitemShut {NoStop}%
\bibitem [{\citenamefont {Herschbach}\ \emph {et~al.}(2012)\citenamefont {Herschbach}, \citenamefont {Pyka}, \citenamefont {Keller},\ and\ \citenamefont {Mehlstäubler}}]{herschbach_linear_2012}%
  \BibitemOpen
  \bibfield  {author} {\bibinfo {author} {\bibfnamefont {N.}~\bibnamefont {Herschbach}}, \bibinfo {author} {\bibfnamefont {K.}~\bibnamefont {Pyka}}, \bibinfo {author} {\bibfnamefont {J.}~\bibnamefont {Keller}},\ and\ \bibinfo {author} {\bibfnamefont {T.~E.}\ \bibnamefont {Mehlstäubler}},\ }\href {https://doi.org/10.1007/s00340-011-4790-y} {\bibfield  {journal} {\bibinfo  {journal} {Applied Physics B}\ }\textbf {\bibinfo {volume} {107}},\ \bibinfo {pages} {891} (\bibinfo {year} {2012})}\BibitemShut {NoStop}%
\bibitem [{\citenamefont {Hausser}\ \emph {et~al.}(2025)\citenamefont {Hausser}, \citenamefont {Keller}, \citenamefont {Nordmann}, \citenamefont {Bhatt}, \citenamefont {Kiethe}, \citenamefont {Liu}, \citenamefont {Richter}, \citenamefont {Von~Boehn}, \citenamefont {Rahm}, \citenamefont {Weyers}, \citenamefont {Benkler}, \citenamefont {Lipphardt}, \citenamefont {Dörscher}, \citenamefont {Stahl}, \citenamefont {Klose}, \citenamefont {Lisdat}, \citenamefont {Filzinger}, \citenamefont {Huntemann}, \citenamefont {Peik},\ and\ \citenamefont {Mehlstäubler}}]{hausser_2025}%
  \BibitemOpen
  \bibfield  {author} {\bibinfo {author} {\bibfnamefont {H.}~\bibnamefont {Hausser}}, \bibinfo {author} {\bibfnamefont {J.}~\bibnamefont {Keller}}, \bibinfo {author} {\bibfnamefont {T.}~\bibnamefont {Nordmann}}, \bibinfo {author} {\bibfnamefont {N.}~\bibnamefont {Bhatt}}, \bibinfo {author} {\bibfnamefont {J.}~\bibnamefont {Kiethe}}, \bibinfo {author} {\bibfnamefont {H.}~\bibnamefont {Liu}}, \bibinfo {author} {\bibfnamefont {I.}~\bibnamefont {Richter}}, \bibinfo {author} {\bibfnamefont {M.}~\bibnamefont {Von~Boehn}}, \bibinfo {author} {\bibfnamefont {J.}~\bibnamefont {Rahm}}, \bibinfo {author} {\bibfnamefont {S.}~\bibnamefont {Weyers}}, \bibinfo {author} {\bibfnamefont {E.}~\bibnamefont {Benkler}}, \bibinfo {author} {\bibfnamefont {B.}~\bibnamefont {Lipphardt}}, \bibinfo {author} {\bibfnamefont {S.}~\bibnamefont {Dörscher}}, \bibinfo {author} {\bibfnamefont {K.}~\bibnamefont {Stahl}}, \bibinfo {author} {\bibfnamefont {J.}~\bibnamefont {Klose}}, \bibinfo {author} {\bibfnamefont {C.}~\bibnamefont {Lisdat}},
  \bibinfo {author} {\bibfnamefont {M.}~\bibnamefont {Filzinger}}, \bibinfo {author} {\bibfnamefont {N.}~\bibnamefont {Huntemann}}, \bibinfo {author} {\bibfnamefont {E.}~\bibnamefont {Peik}},\ and\ \bibinfo {author} {\bibfnamefont {T.}~\bibnamefont {Mehlstäubler}},\ }\href {https://doi.org/10.1103/PhysRevLett.134.023201} {\bibfield  {journal} {\bibinfo  {journal} {Physical Review Letters}\ }\textbf {\bibinfo {volume} {134}},\ \bibinfo {pages} {023201} (\bibinfo {year} {2025})}\BibitemShut {NoStop}%
\bibitem [{\citenamefont {Sokolik}\ \emph {et~al.}(2026)\citenamefont {Sokolik}, \citenamefont {Ozeri},\ and\ \citenamefont {Akerman}}]{sokolik_direct_2026}%
  \BibitemOpen
  \bibfield  {author} {\bibinfo {author} {\bibfnamefont {Y.}~\bibnamefont {Sokolik}}, \bibinfo {author} {\bibfnamefont {R.}~\bibnamefont {Ozeri}},\ and\ \bibinfo {author} {\bibfnamefont {N.}~\bibnamefont {Akerman}},\ }\bibfield  {journal} {\bibinfo  {journal} {arXiv preprint}\ }\href {https://doi.org/10.48550/arXiv.2602.05626} {10.48550/arXiv.2602.05626} (\bibinfo {year} {2026})\BibitemShut {NoStop}%
\bibitem [{\citenamefont {Jordan}\ \emph {et~al.}(2019)\citenamefont {Jordan}, \citenamefont {Gilmore}, \citenamefont {Shankar}, \citenamefont {Safavi-Naini}, \citenamefont {Bohnet}, \citenamefont {Holland},\ and\ \citenamefont {Bollinger}}]{jordan}%
  \BibitemOpen
  \bibfield  {author} {\bibinfo {author} {\bibfnamefont {E.}~\bibnamefont {Jordan}}, \bibinfo {author} {\bibfnamefont {K.~A.}\ \bibnamefont {Gilmore}}, \bibinfo {author} {\bibfnamefont {A.}~\bibnamefont {Shankar}}, \bibinfo {author} {\bibfnamefont {A.}~\bibnamefont {Safavi-Naini}}, \bibinfo {author} {\bibfnamefont {J.~G.}\ \bibnamefont {Bohnet}}, \bibinfo {author} {\bibfnamefont {M.~J.}\ \bibnamefont {Holland}},\ and\ \bibinfo {author} {\bibfnamefont {J.~J.}\ \bibnamefont {Bollinger}},\ }\href {https://doi.org/10.1103/PhysRevLett.122.053603} {\bibfield  {journal} {\bibinfo  {journal} {Physical Review Letters}\ }\textbf {\bibinfo {volume} {122}},\ \bibinfo {pages} {053603} (\bibinfo {year} {2019})}\BibitemShut {NoStop}%
\bibitem [{\citenamefont {Kiesenhofer}\ \emph {et~al.}(2023)\citenamefont {Kiesenhofer}, \citenamefont {Hainzer}, \citenamefont {Zhdanov}, \citenamefont {Holz}, \citenamefont {Bock}, \citenamefont {Ollikainen},\ and\ \citenamefont {Roos}}]{kiesenhofer_controlling_2023}%
  \BibitemOpen
  \bibfield  {author} {\bibinfo {author} {\bibfnamefont {D.}~\bibnamefont {Kiesenhofer}}, \bibinfo {author} {\bibfnamefont {H.}~\bibnamefont {Hainzer}}, \bibinfo {author} {\bibfnamefont {A.}~\bibnamefont {Zhdanov}}, \bibinfo {author} {\bibfnamefont {P.~C.}\ \bibnamefont {Holz}}, \bibinfo {author} {\bibfnamefont {M.}~\bibnamefont {Bock}}, \bibinfo {author} {\bibfnamefont {T.}~\bibnamefont {Ollikainen}},\ and\ \bibinfo {author} {\bibfnamefont {C.~F.}\ \bibnamefont {Roos}},\ }\href {https://doi.org/10.1103/PRXQuantum.4.020317} {\bibfield  {journal} {\bibinfo  {journal} {PRX Quantum}\ }\textbf {\bibinfo {volume} {4}},\ \bibinfo {pages} {020317} (\bibinfo {year} {2023})}\BibitemShut {NoStop}%
\bibitem [{\citenamefont {Jensen}\ \emph {et~al.}(2004)\citenamefont {Jensen}, \citenamefont {Hasegawa},\ and\ \citenamefont {Bollinger}}]{jensen_temperature_2004}%
  \BibitemOpen
  \bibfield  {author} {\bibinfo {author} {\bibfnamefont {M.}~\bibnamefont {Jensen}}, \bibinfo {author} {\bibfnamefont {T.}~\bibnamefont {Hasegawa}},\ and\ \bibinfo {author} {\bibfnamefont {J.}~\bibnamefont {Bollinger}},\ }\href {https://doi.org/10.1103/PhysRevA.70.033401} {\bibfield  {journal} {\bibinfo  {journal} {Physical Review A}\ }\textbf {\bibinfo {volume} {70}},\ \bibinfo {pages} {033401} (\bibinfo {year} {2004})}\BibitemShut {NoStop}%
\bibitem [{\citenamefont {Dubin}\ and\ \citenamefont {O’Neil}(1999)}]{dubin_trapped_1999}%
  \BibitemOpen
  \bibfield  {author} {\bibinfo {author} {\bibfnamefont {D.~H.~E.}\ \bibnamefont {Dubin}}\ and\ \bibinfo {author} {\bibfnamefont {T.~M.}\ \bibnamefont {O’Neil}},\ }\href {https://doi.org/10.1103/RevModPhys.71.87} {\bibfield  {journal} {\bibinfo  {journal} {Reviews of Modern Physics}\ }\textbf {\bibinfo {volume} {71}},\ \bibinfo {pages} {87} (\bibinfo {year} {1999})}\BibitemShut {NoStop}%
\bibitem [{\citenamefont {Vybornyi}\ \emph {et~al.}(2023)\citenamefont {Vybornyi}, \citenamefont {Dreissen}, \citenamefont {Kiesenhofer}, \citenamefont {Hainzer}, \citenamefont {Bock}, \citenamefont {Ollikainen}, \citenamefont {Vadlejch}, \citenamefont {Roos}, \citenamefont {Mehlstäubler},\ and\ \citenamefont {Hammerer}}]{vybornyi_sideband_2023}%
  \BibitemOpen
  \bibfield  {author} {\bibinfo {author} {\bibfnamefont {I.}~\bibnamefont {Vybornyi}}, \bibinfo {author} {\bibfnamefont {L.~S.}\ \bibnamefont {Dreissen}}, \bibinfo {author} {\bibfnamefont {D.}~\bibnamefont {Kiesenhofer}}, \bibinfo {author} {\bibfnamefont {H.}~\bibnamefont {Hainzer}}, \bibinfo {author} {\bibfnamefont {M.}~\bibnamefont {Bock}}, \bibinfo {author} {\bibfnamefont {T.}~\bibnamefont {Ollikainen}}, \bibinfo {author} {\bibfnamefont {D.}~\bibnamefont {Vadlejch}}, \bibinfo {author} {\bibfnamefont {C.~F.}\ \bibnamefont {Roos}}, \bibinfo {author} {\bibfnamefont {T.~E.}\ \bibnamefont {Mehlstäubler}},\ and\ \bibinfo {author} {\bibfnamefont {K.}~\bibnamefont {Hammerer}},\ }\href {https://doi.org/10.1103/PRXQuantum.4.040346} {\bibfield  {journal} {\bibinfo  {journal} {PRX Quantum}\ }\textbf {\bibinfo {volume} {4}},\ \bibinfo {pages} {040346} (\bibinfo {year} {2023})}\BibitemShut {NoStop}%
\bibitem [{\citenamefont {Morigi}\ \emph {et~al.}(2025)\citenamefont {Morigi}, \citenamefont {Bollinger}, \citenamefont {Drewsen}, \citenamefont {Podolsky},\ and\ \citenamefont {Shimshoni}}]{icc_exotic}%
  \BibitemOpen
  \bibfield  {author} {\bibinfo {author} {\bibfnamefont {G.}~\bibnamefont {Morigi}}, \bibinfo {author} {\bibfnamefont {J.}~\bibnamefont {Bollinger}}, \bibinfo {author} {\bibfnamefont {M.}~\bibnamefont {Drewsen}}, \bibinfo {author} {\bibfnamefont {D.}~\bibnamefont {Podolsky}},\ and\ \bibinfo {author} {\bibfnamefont {E.}~\bibnamefont {Shimshoni}},\ }\bibfield  {journal} {\bibinfo  {journal} {arXiv preprint}\ }\href {https://doi.org/10.48550/arXiv.2508.07374} {10.48550/arXiv.2508.07374} (\bibinfo {year} {2025})\BibitemShut {NoStop}%
\bibitem [{\citenamefont {Chen}\ \emph {et~al.}(2020)\citenamefont {Chen}, \citenamefont {Wright}, \citenamefont {Pisenti}, \citenamefont {Murphy}, \citenamefont {Beck}, \citenamefont {Landsman}, \citenamefont {Amini},\ and\ \citenamefont {Nam}}]{chen_efficient-sideband-cooling_2020}%
  \BibitemOpen
  \bibfield  {author} {\bibinfo {author} {\bibfnamefont {J.-S.}\ \bibnamefont {Chen}}, \bibinfo {author} {\bibfnamefont {K.}~\bibnamefont {Wright}}, \bibinfo {author} {\bibfnamefont {N.~C.}\ \bibnamefont {Pisenti}}, \bibinfo {author} {\bibfnamefont {D.}~\bibnamefont {Murphy}}, \bibinfo {author} {\bibfnamefont {K.~M.}\ \bibnamefont {Beck}}, \bibinfo {author} {\bibfnamefont {K.}~\bibnamefont {Landsman}}, \bibinfo {author} {\bibfnamefont {J.~M.}\ \bibnamefont {Amini}},\ and\ \bibinfo {author} {\bibfnamefont {Y.}~\bibnamefont {Nam}},\ }\href {https://doi.org/10.1103/PhysRevA.102.043110} {\bibfield  {journal} {\bibinfo  {journal} {Physical Review A}\ }\textbf {\bibinfo {volume} {102}},\ \bibinfo {pages} {043110} (\bibinfo {year} {2020})}\BibitemShut {NoStop}%
\bibitem [{\citenamefont {Diedrich}\ \emph {et~al.}(1989)\citenamefont {Diedrich}, \citenamefont {Bergquist}, \citenamefont {Itano},\ and\ \citenamefont {Wineland}}]{diedrich_laser_1989}%
  \BibitemOpen
  \bibfield  {author} {\bibinfo {author} {\bibfnamefont {F.}~\bibnamefont {Diedrich}}, \bibinfo {author} {\bibfnamefont {J.~C.}\ \bibnamefont {Bergquist}}, \bibinfo {author} {\bibfnamefont {W.~M.}\ \bibnamefont {Itano}},\ and\ \bibinfo {author} {\bibfnamefont {D.~J.}\ \bibnamefont {Wineland}},\ }\href {https://doi.org/10.1103/PhysRevLett.62.403} {\bibfield  {journal} {\bibinfo  {journal} {Physical Review Letters}\ }\textbf {\bibinfo {volume} {62}},\ \bibinfo {pages} {403} (\bibinfo {year} {1989})}\BibitemShut {NoStop}%
\bibitem [{\citenamefont {King}\ \emph {et~al.}(1998)\citenamefont {King}, \citenamefont {Wood}, \citenamefont {Myatt}, \citenamefont {Turchette}, \citenamefont {Leibfried}, \citenamefont {Itano}, \citenamefont {Monroe},\ and\ \citenamefont {Wineland}}]{king_cooling_1998}%
  \BibitemOpen
  \bibfield  {author} {\bibinfo {author} {\bibfnamefont {B.~E.}\ \bibnamefont {King}}, \bibinfo {author} {\bibfnamefont {C.~S.}\ \bibnamefont {Wood}}, \bibinfo {author} {\bibfnamefont {C.~J.}\ \bibnamefont {Myatt}}, \bibinfo {author} {\bibfnamefont {Q.~A.}\ \bibnamefont {Turchette}}, \bibinfo {author} {\bibfnamefont {D.}~\bibnamefont {Leibfried}}, \bibinfo {author} {\bibfnamefont {W.~M.}\ \bibnamefont {Itano}}, \bibinfo {author} {\bibfnamefont {C.}~\bibnamefont {Monroe}},\ and\ \bibinfo {author} {\bibfnamefont {D.~J.}\ \bibnamefont {Wineland}},\ }\href {https://doi.org/10.1103/PhysRevLett.81.1525} {\bibfield  {journal} {\bibinfo  {journal} {Physical Review Letters}\ }\textbf {\bibinfo {volume} {81}},\ \bibinfo {pages} {1525} (\bibinfo {year} {1998})}\BibitemShut {NoStop}%
\bibitem [{\citenamefont {Monroe}\ \emph {et~al.}(1995)\citenamefont {Monroe}, \citenamefont {Meekhof}, \citenamefont {King}, \citenamefont {Jefferts}, \citenamefont {Itano}, \citenamefont {Wineland},\ and\ \citenamefont {Gould}}]{monroe_resolved_sideband_1995}%
  \BibitemOpen
  \bibfield  {author} {\bibinfo {author} {\bibfnamefont {C.}~\bibnamefont {Monroe}}, \bibinfo {author} {\bibfnamefont {D.~M.}\ \bibnamefont {Meekhof}}, \bibinfo {author} {\bibfnamefont {B.~E.}\ \bibnamefont {King}}, \bibinfo {author} {\bibfnamefont {S.~R.}\ \bibnamefont {Jefferts}}, \bibinfo {author} {\bibfnamefont {W.~M.}\ \bibnamefont {Itano}}, \bibinfo {author} {\bibfnamefont {D.~J.}\ \bibnamefont {Wineland}},\ and\ \bibinfo {author} {\bibfnamefont {P.}~\bibnamefont {Gould}},\ }\href {https://doi.org/10.1103/PhysRevLett.75.4011} {\bibfield  {journal} {\bibinfo  {journal} {Physical Review Letters}\ }\textbf {\bibinfo {volume} {75}},\ \bibinfo {pages} {4011} (\bibinfo {year} {1995})}\BibitemShut {NoStop}%
\bibitem [{\citenamefont {Wu}\ \emph {et~al.}(2023)\citenamefont {Wu}, \citenamefont {Shi},\ and\ \citenamefont {Zhang}}]{wu_continuous_2023}%
  \BibitemOpen
  \bibfield  {author} {\bibinfo {author} {\bibfnamefont {Q.}~\bibnamefont {Wu}}, \bibinfo {author} {\bibfnamefont {Y.}~\bibnamefont {Shi}},\ and\ \bibinfo {author} {\bibfnamefont {J.}~\bibnamefont {Zhang}},\ }\href {https://doi.org/10.1103/PhysRevResearch.5.023022} {\bibfield  {journal} {\bibinfo  {journal} {Physical Review Research}\ }\textbf {\bibinfo {volume} {5}},\ \bibinfo {pages} {023022} (\bibinfo {year} {2023})}\BibitemShut {NoStop}%
\bibitem [{\citenamefont {Morigi}\ \emph {et~al.}(2000)\citenamefont {Morigi}, \citenamefont {Eschner},\ and\ \citenamefont {Keitel}}]{morigi_ground_2000}%
  \BibitemOpen
  \bibfield  {author} {\bibinfo {author} {\bibfnamefont {G.}~\bibnamefont {Morigi}}, \bibinfo {author} {\bibfnamefont {J.}~\bibnamefont {Eschner}},\ and\ \bibinfo {author} {\bibfnamefont {C.~H.}\ \bibnamefont {Keitel}},\ }\href {https://doi.org/10.1103/PhysRevLett.85.4458} {\bibfield  {journal} {\bibinfo  {journal} {Physical Review Letters}\ }\textbf {\bibinfo {volume} {85}},\ \bibinfo {pages} {4458} (\bibinfo {year} {2000})}\BibitemShut {NoStop}%
\bibitem [{\citenamefont {Morigi}(2003{\natexlab{a}})}]{morigi_cooling_2003}%
  \BibitemOpen
  \bibfield  {author} {\bibinfo {author} {\bibfnamefont {G.}~\bibnamefont {Morigi}},\ }\href {https://doi.org/10.1103/PhysRevA.67.033402} {\bibfield  {journal} {\bibinfo  {journal} {Physical Review A}\ }\textbf {\bibinfo {volume} {67}},\ \bibinfo {pages} {033402} (\bibinfo {year} {2003}{\natexlab{a}})}\BibitemShut {NoStop}%
\bibitem [{\citenamefont {Roos}\ \emph {et~al.}(2000)\citenamefont {Roos}, \citenamefont {Leibfried}, \citenamefont {Mundt}, \citenamefont {Schmidt-Kaler}, \citenamefont {Eschner},\ and\ \citenamefont {Blatt}}]{roos_experimental_2000}%
  \BibitemOpen
  \bibfield  {author} {\bibinfo {author} {\bibfnamefont {C.~F.}\ \bibnamefont {Roos}}, \bibinfo {author} {\bibfnamefont {D.}~\bibnamefont {Leibfried}}, \bibinfo {author} {\bibfnamefont {A.}~\bibnamefont {Mundt}}, \bibinfo {author} {\bibfnamefont {F.}~\bibnamefont {Schmidt-Kaler}}, \bibinfo {author} {\bibfnamefont {J.}~\bibnamefont {Eschner}},\ and\ \bibinfo {author} {\bibfnamefont {R.}~\bibnamefont {Blatt}},\ }\href {https://doi.org/10.1103/PhysRevLett.85.5547} {\bibfield  {journal} {\bibinfo  {journal} {Physical Review Letters}\ }\textbf {\bibinfo {volume} {85}},\ \bibinfo {pages} {5547} (\bibinfo {year} {2000})}\BibitemShut {NoStop}%
\bibitem [{\citenamefont {Zhang}\ \emph {et~al.}(2022)\citenamefont {Zhang}, \citenamefont {Zhang}, \citenamefont {Xie}, \citenamefont {Wu}, \citenamefont {Ou}, \citenamefont {Chen}, \citenamefont {Bao}, \citenamefont {Haljan}, \citenamefont {Wu}, \citenamefont {Zhang},\ and\ \citenamefont {Chen}}]{zhang_eit}%
  \BibitemOpen
  \bibfield  {author} {\bibinfo {author} {\bibfnamefont {J.}~\bibnamefont {Zhang}}, \bibinfo {author} {\bibfnamefont {M.-C.}\ \bibnamefont {Zhang}}, \bibinfo {author} {\bibfnamefont {Y.}~\bibnamefont {Xie}}, \bibinfo {author} {\bibfnamefont {C.-W.}\ \bibnamefont {Wu}}, \bibinfo {author} {\bibfnamefont {B.-Q.}\ \bibnamefont {Ou}}, \bibinfo {author} {\bibfnamefont {T.}~\bibnamefont {Chen}}, \bibinfo {author} {\bibfnamefont {W.-S.}\ \bibnamefont {Bao}}, \bibinfo {author} {\bibfnamefont {P.}~\bibnamefont {Haljan}}, \bibinfo {author} {\bibfnamefont {W.}~\bibnamefont {Wu}}, \bibinfo {author} {\bibfnamefont {S.}~\bibnamefont {Zhang}},\ and\ \bibinfo {author} {\bibfnamefont {P.-X.}\ \bibnamefont {Chen}},\ }\href {https://doi.org/10.1103/PhysRevApplied.18.014022} {\bibfield  {journal} {\bibinfo  {journal} {Physical Review Applied}\ }\textbf {\bibinfo {volume} {18}},\ \bibinfo {pages} {014022} (\bibinfo {year} {2022})}\BibitemShut {NoStop}%
\bibitem [{\citenamefont {Qiao}\ \emph {et~al.}(2021)\citenamefont {Qiao}, \citenamefont {Wang}, \citenamefont {Cai}, \citenamefont {Du}, \citenamefont {Wang}, \citenamefont {Luan}, \citenamefont {Chen}, \citenamefont {Noh},\ and\ \citenamefont {Kim}}]{qiao_eit}%
  \BibitemOpen
  \bibfield  {author} {\bibinfo {author} {\bibfnamefont {M.}~\bibnamefont {Qiao}}, \bibinfo {author} {\bibfnamefont {Y.}~\bibnamefont {Wang}}, \bibinfo {author} {\bibfnamefont {Z.}~\bibnamefont {Cai}}, \bibinfo {author} {\bibfnamefont {B.}~\bibnamefont {Du}}, \bibinfo {author} {\bibfnamefont {P.}~\bibnamefont {Wang}}, \bibinfo {author} {\bibfnamefont {C.}~\bibnamefont {Luan}}, \bibinfo {author} {\bibfnamefont {W.}~\bibnamefont {Chen}}, \bibinfo {author} {\bibfnamefont {H.-R.}\ \bibnamefont {Noh}},\ and\ \bibinfo {author} {\bibfnamefont {K.}~\bibnamefont {Kim}},\ }\href {https://doi.org/10.1103/PhysRevLett.126.023604} {\bibfield  {journal} {\bibinfo  {journal} {Physical Review Letters}\ }\textbf {\bibinfo {volume} {126}},\ \bibinfo {pages} {023604} (\bibinfo {year} {2021})}\BibitemShut {NoStop}%
\bibitem [{\citenamefont {Feng}\ \emph {et~al.}(2020)\citenamefont {Feng}, \citenamefont {Tan}, \citenamefont {De}, \citenamefont {Menon}, \citenamefont {Chu}, \citenamefont {Pagano},\ and\ \citenamefont {Monroe}}]{monroe}%
  \BibitemOpen
  \bibfield  {author} {\bibinfo {author} {\bibfnamefont {L.}~\bibnamefont {Feng}}, \bibinfo {author} {\bibfnamefont {W.}~\bibnamefont {Tan}}, \bibinfo {author} {\bibfnamefont {A.}~\bibnamefont {De}}, \bibinfo {author} {\bibfnamefont {A.}~\bibnamefont {Menon}}, \bibinfo {author} {\bibfnamefont {A.}~\bibnamefont {Chu}}, \bibinfo {author} {\bibfnamefont {G.}~\bibnamefont {Pagano}},\ and\ \bibinfo {author} {\bibfnamefont {C.}~\bibnamefont {Monroe}},\ }\href {https://doi.org/10.1103/PhysRevLett.125.053001} {\bibfield  {journal} {\bibinfo  {journal} {Physical Review Letters}\ }\textbf {\bibinfo {volume} {125}},\ \bibinfo {pages} {053001} (\bibinfo {year} {2020})}\BibitemShut {NoStop}%
\bibitem [{\citenamefont {Khan}\ \emph {et~al.}(2026)\citenamefont {Khan}, \citenamefont {Wellnitz}, \citenamefont {Sundar}, \citenamefont {Zhang}, \citenamefont {Carter}, \citenamefont {Bollinger}, \citenamefont {Shankar},\ and\ \citenamefont {Rey}}]{khan_many_body_2026}%
  \BibitemOpen
  \bibfield  {author} {\bibinfo {author} {\bibfnamefont {M.~M.}\ \bibnamefont {Khan}}, \bibinfo {author} {\bibfnamefont {D.}~\bibnamefont {Wellnitz}}, \bibinfo {author} {\bibfnamefont {B.}~\bibnamefont {Sundar}}, \bibinfo {author} {\bibfnamefont {H.}~\bibnamefont {Zhang}}, \bibinfo {author} {\bibfnamefont {A.}~\bibnamefont {Carter}}, \bibinfo {author} {\bibfnamefont {J.~J.}\ \bibnamefont {Bollinger}}, \bibinfo {author} {\bibfnamefont {A.}~\bibnamefont {Shankar}},\ and\ \bibinfo {author} {\bibfnamefont {A.~M.}\ \bibnamefont {Rey}},\ }\bibfield  {journal} {\bibinfo  {journal} {arXiv preprint}\ }\href {https://doi.org/10.48550/arXiv.2601.09180} {10.48550/arXiv.2601.09180} (\bibinfo {year} {2026}),\ \bibinfo {note} {arXiv:2601.09180}\BibitemShut {NoStop}%
\bibitem [{\citenamefont {Wan}\ \emph {et~al.}(2015)\citenamefont {Wan}, \citenamefont {Gebert}, \citenamefont {Wolf},\ and\ \citenamefont {Schmidt}}]{wan_efficient_2015}%
  \BibitemOpen
  \bibfield  {author} {\bibinfo {author} {\bibfnamefont {Y.}~\bibnamefont {Wan}}, \bibinfo {author} {\bibfnamefont {F.}~\bibnamefont {Gebert}}, \bibinfo {author} {\bibfnamefont {F.}~\bibnamefont {Wolf}},\ and\ \bibinfo {author} {\bibfnamefont {P.~O.}\ \bibnamefont {Schmidt}},\ }\href {https://doi.org/10.1103/PhysRevA.91.043425} {\bibfield  {journal} {\bibinfo  {journal} {Physical Review A}\ }\textbf {\bibinfo {volume} {91}},\ \bibinfo {pages} {043425} (\bibinfo {year} {2015})}\BibitemShut {NoStop}%
\bibitem [{\citenamefont {Rasmusson}\ \emph {et~al.}(2021)\citenamefont {Rasmusson}, \citenamefont {D'Onofrio}, \citenamefont {Xie}, \citenamefont {Cui},\ and\ \citenamefont {Richerme}}]{rasmusson_optimized_2021}%
  \BibitemOpen
  \bibfield  {author} {\bibinfo {author} {\bibfnamefont {A.~J.}\ \bibnamefont {Rasmusson}}, \bibinfo {author} {\bibfnamefont {M.}~\bibnamefont {D'Onofrio}}, \bibinfo {author} {\bibfnamefont {Y.}~\bibnamefont {Xie}}, \bibinfo {author} {\bibfnamefont {J.}~\bibnamefont {Cui}},\ and\ \bibinfo {author} {\bibfnamefont {P.}~\bibnamefont {Richerme}},\ }\href {https://doi.org/10.1103/PhysRevA.104.043108} {\bibfield  {journal} {\bibinfo  {journal} {Physical Review A}\ }\textbf {\bibinfo {volume} {104}},\ \bibinfo {pages} {043108} (\bibinfo {year} {2021})}\BibitemShut {NoStop}%
\bibitem [{\citenamefont {Reed}\ \emph {et~al.}(2024)\citenamefont {Reed}, \citenamefont {Qi},\ and\ \citenamefont {Brown}}]{reed_comparison_2024}%
  \BibitemOpen
  \bibfield  {author} {\bibinfo {author} {\bibfnamefont {E.~C.}\ \bibnamefont {Reed}}, \bibinfo {author} {\bibfnamefont {L.}~\bibnamefont {Qi}},\ and\ \bibinfo {author} {\bibfnamefont {K.~R.}\ \bibnamefont {Brown}},\ }\href {https://doi.org/10.1103/PhysRevA.110.013123} {\bibfield  {journal} {\bibinfo  {journal} {Physical Review A}\ }\textbf {\bibinfo {volume} {110}},\ \bibinfo {pages} {013123} (\bibinfo {year} {2024})}\BibitemShut {NoStop}%
\bibitem [{\citenamefont {Stutter}\ \emph {et~al.}(2018)\citenamefont {Stutter}, \citenamefont {Hrmo}, \citenamefont {Jarlaud}, \citenamefont {Joshi}, \citenamefont {Goodwin},\ and\ \citenamefont {Thompson}}]{stutter_sideband_2018}%
  \BibitemOpen
  \bibfield  {author} {\bibinfo {author} {\bibfnamefont {G.}~\bibnamefont {Stutter}}, \bibinfo {author} {\bibfnamefont {P.}~\bibnamefont {Hrmo}}, \bibinfo {author} {\bibfnamefont {V.}~\bibnamefont {Jarlaud}}, \bibinfo {author} {\bibfnamefont {M.~K.}\ \bibnamefont {Joshi}}, \bibinfo {author} {\bibfnamefont {J.~F.}\ \bibnamefont {Goodwin}},\ and\ \bibinfo {author} {\bibfnamefont {R.~C.}\ \bibnamefont {Thompson}},\ }\href {https://doi.org/10.1080/09500340.2017.1376719} {\bibfield  {journal} {\bibinfo  {journal} {Journal of Modern Optics}\ }\textbf {\bibinfo {volume} {65}},\ \bibinfo {pages} {549} (\bibinfo {year} {2018})}\BibitemShut {NoStop}%
\bibitem [{\citenamefont {Rasmusson}(2026)}]{rasmusson_high_2026}%
  \BibitemOpen
  \bibfield  {author} {\bibinfo {author} {\bibfnamefont {A.~J.}\ \bibnamefont {Rasmusson}},\ }\bibfield  {journal} {\bibinfo  {journal} {arXiv preprint}\ }\href {https://doi.org/10.48550/arXiv.2604.03435} {10.48550/arXiv.2604.03435} (\bibinfo {year} {2026})\BibitemShut {NoStop}%
\bibitem [{\citenamefont {Chen}\ \emph {et~al.}(2017)\citenamefont {Chen}, \citenamefont {Brewer}, \citenamefont {Chou}, \citenamefont {Wineland}, \citenamefont {Leibrandt},\ and\ \citenamefont {Hume}}]{chen_sympathetic_2017}%
  \BibitemOpen
  \bibfield  {author} {\bibinfo {author} {\bibfnamefont {J.-S.}\ \bibnamefont {Chen}}, \bibinfo {author} {\bibfnamefont {S.}~\bibnamefont {Brewer}}, \bibinfo {author} {\bibfnamefont {C.}~\bibnamefont {Chou}}, \bibinfo {author} {\bibfnamefont {D.}~\bibnamefont {Wineland}}, \bibinfo {author} {\bibfnamefont {D.}~\bibnamefont {Leibrandt}},\ and\ \bibinfo {author} {\bibfnamefont {D.}~\bibnamefont {Hume}},\ }\href {https://doi.org/10.1103/PhysRevLett.118.053002} {\bibfield  {journal} {\bibinfo  {journal} {Physical Review Letters}\ }\textbf {\bibinfo {volume} {118}},\ \bibinfo {pages} {053002} (\bibinfo {year} {2017})}\BibitemShut {NoStop}%
\bibitem [{\citenamefont {Hankin}\ \emph {et~al.}(2019)\citenamefont {Hankin}, \citenamefont {Clements}, \citenamefont {Huang}, \citenamefont {Brewer}, \citenamefont {Chen}, \citenamefont {Chou}, \citenamefont {Hume},\ and\ \citenamefont {Leibrandt}}]{hankin_systematic_2019}%
  \BibitemOpen
  \bibfield  {author} {\bibinfo {author} {\bibfnamefont {A.~M.}\ \bibnamefont {Hankin}}, \bibinfo {author} {\bibfnamefont {E.~R.}\ \bibnamefont {Clements}}, \bibinfo {author} {\bibfnamefont {Y.}~\bibnamefont {Huang}}, \bibinfo {author} {\bibfnamefont {S.~M.}\ \bibnamefont {Brewer}}, \bibinfo {author} {\bibfnamefont {J.-S.}\ \bibnamefont {Chen}}, \bibinfo {author} {\bibfnamefont {C.~W.}\ \bibnamefont {Chou}}, \bibinfo {author} {\bibfnamefont {D.~B.}\ \bibnamefont {Hume}},\ and\ \bibinfo {author} {\bibfnamefont {D.~R.}\ \bibnamefont {Leibrandt}},\ }\href {https://doi.org/10.1103/PhysRevA.100.033419} {\bibfield  {journal} {\bibinfo  {journal} {Physical Review A}\ }\textbf {\bibinfo {volume} {100}},\ \bibinfo {pages} {033419} (\bibinfo {year} {2019})}\BibitemShut {NoStop}%
\bibitem [{\citenamefont {Shankar}\ \emph {et~al.}(2019)\citenamefont {Shankar}, \citenamefont {Jordan}, \citenamefont {Gilmore}, \citenamefont {Safavi-Naini}, \citenamefont {Bollinger},\ and\ \citenamefont {Holland}}]{shankar}%
  \BibitemOpen
  \bibfield  {author} {\bibinfo {author} {\bibfnamefont {A.}~\bibnamefont {Shankar}}, \bibinfo {author} {\bibfnamefont {E.}~\bibnamefont {Jordan}}, \bibinfo {author} {\bibfnamefont {K.~A.}\ \bibnamefont {Gilmore}}, \bibinfo {author} {\bibfnamefont {A.}~\bibnamefont {Safavi-Naini}}, \bibinfo {author} {\bibfnamefont {J.~J.}\ \bibnamefont {Bollinger}},\ and\ \bibinfo {author} {\bibfnamefont {M.~J.}\ \bibnamefont {Holland}},\ }\href {https://doi.org/10.1103/PhysRevA.99.023409} {\bibfield  {journal} {\bibinfo  {journal} {Physical Review A}\ }\textbf {\bibinfo {volume} {99}},\ \bibinfo {pages} {023409} (\bibinfo {year} {2019})}\BibitemShut {NoStop}%
\bibitem [{sup(2026)}]{supplemental}%
  \BibitemOpen
  \href@noop {} {\bibinfo {title} {See supplemental material for derivations and theoretical details.}} (\bibinfo {year} {2026})\BibitemShut {NoStop}%
\bibitem [{\citenamefont {Leibfried}\ \emph {et~al.}(2003{\natexlab{a}})\citenamefont {Leibfried}, \citenamefont {Blatt}, \citenamefont {Monroe},\ and\ \citenamefont {Wineland}}]{leibfried_quantum_2003}%
  \BibitemOpen
  \bibfield  {author} {\bibinfo {author} {\bibfnamefont {D.}~\bibnamefont {Leibfried}}, \bibinfo {author} {\bibfnamefont {R.}~\bibnamefont {Blatt}}, \bibinfo {author} {\bibfnamefont {C.}~\bibnamefont {Monroe}},\ and\ \bibinfo {author} {\bibfnamefont {D.}~\bibnamefont {Wineland}},\ }\href {https://doi.org/10.1103/RevModPhys.75.281} {\bibfield  {journal} {\bibinfo  {journal} {Reviews of Modern Physics}\ }\textbf {\bibinfo {volume} {75}},\ \bibinfo {pages} {281} (\bibinfo {year} {2003}{\natexlab{a}})}\BibitemShut {NoStop}%
\bibitem [{\citenamefont {Leibfried}\ \emph {et~al.}(2003{\natexlab{b}})\citenamefont {Leibfried}, \citenamefont {Blatt}, \citenamefont {Monroe},\ and\ \citenamefont {Wineland}}]{Leibfried2003}%
  \BibitemOpen
  \bibfield  {author} {\bibinfo {author} {\bibfnamefont {D.}~\bibnamefont {Leibfried}}, \bibinfo {author} {\bibfnamefont {R.}~\bibnamefont {Blatt}}, \bibinfo {author} {\bibfnamefont {C.}~\bibnamefont {Monroe}},\ and\ \bibinfo {author} {\bibfnamefont {D.}~\bibnamefont {Wineland}},\ }\href {https://doi.org/10.1103/RevModPhys.75.281} {\bibfield  {journal} {\bibinfo  {journal} {Rev. Mod. Phys.}\ }\textbf {\bibinfo {volume} {75}},\ \bibinfo {pages} {281} (\bibinfo {year} {2003}{\natexlab{b}})}\BibitemShut {NoStop}%
\bibitem [{\citenamefont {Hume}\ \emph {et~al.}(2009)\citenamefont {Hume}, \citenamefont {Chou}, \citenamefont {Rosenband},\ and\ \citenamefont {Wineland}}]{hume_preparation_2009}%
  \BibitemOpen
  \bibfield  {author} {\bibinfo {author} {\bibfnamefont {D.~B.}\ \bibnamefont {Hume}}, \bibinfo {author} {\bibfnamefont {C.~W.}\ \bibnamefont {Chou}}, \bibinfo {author} {\bibfnamefont {T.}~\bibnamefont {Rosenband}},\ and\ \bibinfo {author} {\bibfnamefont {D.~J.}\ \bibnamefont {Wineland}},\ }\href {https://doi.org/10.1103/PhysRevA.80.052302} {\bibfield  {journal} {\bibinfo  {journal} {Physical Review A}\ }\textbf {\bibinfo {volume} {80}},\ \bibinfo {pages} {052302} (\bibinfo {year} {2009})}\BibitemShut {NoStop}%
\bibitem [{\citenamefont {Lechner}\ \emph {et~al.}(2016)\citenamefont {Lechner}, \citenamefont {Maier}, \citenamefont {Hempel}, \citenamefont {Jurcevic}, \citenamefont {Lanyon}, \citenamefont {Monz}, \citenamefont {Brownnutt}, \citenamefont {Blatt},\ and\ \citenamefont {Roos}}]{Lechner2016}%
  \BibitemOpen
  \bibfield  {author} {\bibinfo {author} {\bibfnamefont {R.}~\bibnamefont {Lechner}}, \bibinfo {author} {\bibfnamefont {C.}~\bibnamefont {Maier}}, \bibinfo {author} {\bibfnamefont {C.}~\bibnamefont {Hempel}}, \bibinfo {author} {\bibfnamefont {P.}~\bibnamefont {Jurcevic}}, \bibinfo {author} {\bibfnamefont {B.~P.}\ \bibnamefont {Lanyon}}, \bibinfo {author} {\bibfnamefont {T.}~\bibnamefont {Monz}}, \bibinfo {author} {\bibfnamefont {M.}~\bibnamefont {Brownnutt}}, \bibinfo {author} {\bibfnamefont {R.}~\bibnamefont {Blatt}},\ and\ \bibinfo {author} {\bibfnamefont {C.~F.}\ \bibnamefont {Roos}},\ }\href {https://doi.org/10.1103/PhysRevA.93.053401} {\bibfield  {journal} {\bibinfo  {journal} {Phys. Rev. A}\ }\textbf {\bibinfo {volume} {93}},\ \bibinfo {pages} {053401} (\bibinfo {year} {2016})}\BibitemShut {NoStop}%
\bibitem [{\citenamefont {Kirkova}\ \emph {et~al.}(2021)\citenamefont {Kirkova}, \citenamefont {Li},\ and\ \citenamefont {Ivanov}}]{kirkova_adiabatic_2021}%
  \BibitemOpen
  \bibfield  {author} {\bibinfo {author} {\bibfnamefont {A.~V.}\ \bibnamefont {Kirkova}}, \bibinfo {author} {\bibfnamefont {W.}~\bibnamefont {Li}},\ and\ \bibinfo {author} {\bibfnamefont {P.~A.}\ \bibnamefont {Ivanov}},\ }\href {https://doi.org/10.1103/PhysRevResearch.3.013244} {\bibfield  {journal} {\bibinfo  {journal} {Phys. Rev. Res.}\ }\textbf {\bibinfo {volume} {3}},\ \bibinfo {pages} {013244} (\bibinfo {year} {2021})}\BibitemShut {NoStop}%
\bibitem [{\citenamefont {Rajakumar}\ \emph {et~al.}(2026)\citenamefont {Rajakumar}, \citenamefont {Pandey}, \citenamefont {Das},\ and\ \citenamefont {Shankar}}]{shankar_new}%
  \BibitemOpen
  \bibfield  {author} {\bibinfo {author} {\bibfnamefont {K.}~\bibnamefont {Rajakumar}}, \bibinfo {author} {\bibfnamefont {A.}~\bibnamefont {Pandey}}, \bibinfo {author} {\bibfnamefont {A.}~\bibnamefont {Das}},\ and\ \bibinfo {author} {\bibfnamefont {A.}~\bibnamefont {Shankar}},\ }\href@noop {} {\bibinfo {title} {Quantum beam-splitter cooling and thermometry in large trapped-ion crystals}} (\bibinfo {year} {2026}),\ \bibinfo {note} {forthcoming preprint}\BibitemShut {NoStop}%
\bibitem [{\citenamefont {Morigi}(2003{\natexlab{b}})}]{morigi}%
  \BibitemOpen
  \bibfield  {author} {\bibinfo {author} {\bibfnamefont {G.}~\bibnamefont {Morigi}},\ }\href {https://doi.org/10.1103/PhysRevA.67.033402} {\bibfield  {journal} {\bibinfo  {journal} {Physical Review A}\ }\textbf {\bibinfo {volume} {67}},\ \bibinfo {pages} {033402} (\bibinfo {year} {2003}{\natexlab{b}})}\BibitemShut {NoStop}%
\bibitem [{\citenamefont {Varshalovich}\ \emph {et~al.}(1988)\citenamefont {Varshalovich}, \citenamefont {Moskalev},\ and\ \citenamefont {Khersonskii}}]{varshalovich}%
  \BibitemOpen
  \bibfield  {author} {\bibinfo {author} {\bibfnamefont {D.~A.}\ \bibnamefont {Varshalovich}}, \bibinfo {author} {\bibfnamefont {A.~N.}\ \bibnamefont {Moskalev}},\ and\ \bibinfo {author} {\bibfnamefont {V.~K.}\ \bibnamefont {Khersonskii}},\ }\href {https://doi.org/10.1142/0270} {\emph {\bibinfo {title} {Quantum {Theory} of {Angular} {Momentum}}}}\ (\bibinfo  {publisher} {World Scientific},\ \bibinfo {year} {1988})\BibitemShut {NoStop}%
\bibitem [{\citenamefont {Kerber}\ \emph {et~al.}(2025)\citenamefont {Kerber}, \citenamefont {Ritsch},\ and\ \citenamefont {Ostermann}}]{kerber_cumulants_2025}%
  \BibitemOpen
  \bibfield  {author} {\bibinfo {author} {\bibfnamefont {J.}~\bibnamefont {Kerber}}, \bibinfo {author} {\bibfnamefont {H.}~\bibnamefont {Ritsch}},\ and\ \bibinfo {author} {\bibfnamefont {L.}~\bibnamefont {Ostermann}},\ }\bibfield  {journal} {\bibinfo  {journal} {arXiv preprint}\ }\href {https://doi.org/10.48550/arXiv.2511.20115} {10.48550/arXiv.2511.20115} (\bibinfo {year} {2025})\BibitemShut {NoStop}%
\bibitem [{\citenamefont {Kirton}\ \emph {et~al.}(2019)\citenamefont {Kirton}, \citenamefont {Roses}, \citenamefont {Keeling},\ and\ \citenamefont {Dalla~Torre}}]{kirton_introduction_2019}%
  \BibitemOpen
  \bibfield  {author} {\bibinfo {author} {\bibfnamefont {P.}~\bibnamefont {Kirton}}, \bibinfo {author} {\bibfnamefont {M.~M.}\ \bibnamefont {Roses}}, \bibinfo {author} {\bibfnamefont {J.}~\bibnamefont {Keeling}},\ and\ \bibinfo {author} {\bibfnamefont {E.~G.}\ \bibnamefont {Dalla~Torre}},\ }\href {https://doi.org/10.1002/qute.201800043} {\bibfield  {journal} {\bibinfo  {journal} {Advanced Quantum Technologies}\ }\textbf {\bibinfo {volume} {2}},\ \bibinfo {pages} {1800043} (\bibinfo {year} {2019})}\BibitemShut {NoStop}%
\bibitem [{\citenamefont {Piñeiro~Orioli}\ \emph {et~al.}(2017)\citenamefont {Piñeiro~Orioli}, \citenamefont {Safavi-Naini}, \citenamefont {Wall},\ and\ \citenamefont {Rey}}]{pineiro_orioli_nonequilibrium_2017}%
  \BibitemOpen
  \bibfield  {author} {\bibinfo {author} {\bibfnamefont {A.}~\bibnamefont {Piñeiro~Orioli}}, \bibinfo {author} {\bibfnamefont {A.}~\bibnamefont {Safavi-Naini}}, \bibinfo {author} {\bibfnamefont {M.~L.}\ \bibnamefont {Wall}},\ and\ \bibinfo {author} {\bibfnamefont {A.~M.}\ \bibnamefont {Rey}},\ }\href {https://doi.org/10.1103/PhysRevA.96.033607} {\bibfield  {journal} {\bibinfo  {journal} {Physical Review A}\ }\textbf {\bibinfo {volume} {96}},\ \bibinfo {pages} {033607} (\bibinfo {year} {2017})}\BibitemShut {NoStop}%
\bibitem [{\citenamefont {Schachenmayer}\ \emph {et~al.}(2015)\citenamefont {Schachenmayer}, \citenamefont {Pikovski},\ and\ \citenamefont {Rey}}]{schachenmayer_many-body_2015}%
  \BibitemOpen
  \bibfield  {author} {\bibinfo {author} {\bibfnamefont {J.}~\bibnamefont {Schachenmayer}}, \bibinfo {author} {\bibfnamefont {A.}~\bibnamefont {Pikovski}},\ and\ \bibinfo {author} {\bibfnamefont {A.}~\bibnamefont {Rey}},\ }\href {https://doi.org/10.1103/PhysRevX.5.011022} {\bibfield  {journal} {\bibinfo  {journal} {Physical Review X}\ }\textbf {\bibinfo {volume} {5}},\ \bibinfo {pages} {011022} (\bibinfo {year} {2015})}\BibitemShut {NoStop}%
\bibitem [{\citenamefont {Wootters}(1987)}]{wootters_wigner-function_1987}%
  \BibitemOpen
  \bibfield  {author} {\bibinfo {author} {\bibfnamefont {W.~K.}\ \bibnamefont {Wootters}},\ }\href {https://doi.org/10.1016/0003-4916(87)90176-X} {\bibfield  {journal} {\bibinfo  {journal} {Annals of Physics}\ }\textbf {\bibinfo {volume} {176}},\ \bibinfo {pages} {1} (\bibinfo {year} {1987})}\BibitemShut {NoStop}%
\bibitem [{\citenamefont {Huber}\ \emph {et~al.}(2022)\citenamefont {Huber}, \citenamefont {Rey},\ and\ \citenamefont {Rabl}}]{huber_realistic_2022}%
  \BibitemOpen
  \bibfield  {author} {\bibinfo {author} {\bibfnamefont {J.}~\bibnamefont {Huber}}, \bibinfo {author} {\bibfnamefont {A.~M.}\ \bibnamefont {Rey}},\ and\ \bibinfo {author} {\bibfnamefont {P.}~\bibnamefont {Rabl}},\ }\href {https://doi.org/10.1103/PhysRevA.105.013716} {\bibfield  {journal} {\bibinfo  {journal} {Physical Review A}\ }\textbf {\bibinfo {volume} {105}},\ \bibinfo {pages} {013716} (\bibinfo {year} {2022})}\BibitemShut {NoStop}%
\bibitem [{\citenamefont {Hosseinabadi}\ \emph {et~al.}(2025)\citenamefont {Hosseinabadi}, \citenamefont {Chelpanova},\ and\ \citenamefont {Marino}}]{hosseinabadi_user_friendly_2025}%
  \BibitemOpen
  \bibfield  {author} {\bibinfo {author} {\bibfnamefont {H.}~\bibnamefont {Hosseinabadi}}, \bibinfo {author} {\bibfnamefont {O.}~\bibnamefont {Chelpanova}},\ and\ \bibinfo {author} {\bibfnamefont {J.}~\bibnamefont {Marino}},\ }\href {https://doi.org/10.1103/1wwv-k7hg} {\bibfield  {journal} {\bibinfo  {journal} {PRX Quantum}\ }\textbf {\bibinfo {volume} {6}},\ \bibinfo {pages} {030344} (\bibinfo {year} {2025})}\BibitemShut {NoStop}%
\end{thebibliography}
\end{document}